\documentclass[%
 a4paper,reprint,twocolumn,
 longbibliography,
 notitlepage,showkeys,
superscriptaddress,nofootinbib,
 amsmath,amssymb,
 aps,
prd,
floatfix,
]{revtex4-2}
\usepackage{epsfig}
\usepackage{dsfont}
\usepackage[x11names]{xcolor}
\usepackage{hyperref}
\usepackage{orcidlink}
\hypersetup{           
    colorlinks=true,                
    breaklinks=true,                
    urlcolor= Purple2,                
    linkcolor= red,                
    bookmarksopen=false,
    filecolor=black,
    citecolor= magenta,
    linkbordercolor=blue}

\usepackage[T1]{fontenc}
\usepackage{graphicx}
\usepackage{orcidlink}
\begin{document}
\title{Frolov Black Hole Surrounded by Quintessence - I: Thermodynamics, Geodesics and Shadows}

\author{Mrinnoy M. Gohain\orcidlink{0000-0002-1097-2124}}
\email{mrinmoygohain19@gmail.com}
\affiliation{%
 Department of Physics, Dibrugarh University, Dibrugarh \\
 Assam, India, 786004}

\author{Kalyan Bhuyan\orcidlink{0000-0002-8896-7691}}%
 \email{kalyanbhuyan@dibru.ac.in}
\affiliation{%
 Department of Physics, Dibrugarh University, Dibrugarh \\
 Assam, India, 786004}%
 \affiliation{Theoretical Physics Divison, Centre for Atmospheric Studies, Dibrugarh University, Dibrugarh, Assam, India 786004}

%\author{Hari Prasad Saikia \orcidlink{0009-0008-9048-7719}}%
% \email{hariprasadsaikia@dibru.ac.in}
%\affiliation{%
% Department of Physics, Dibrugarh University, Dibrugarh \\
% Assam, India, 786004}
\author{Rajnandini Borgohain \orcidlink{0009-0005-6029-9424}}%
 \email{rajnandiniborgohainnzr@gmail.com}
\affiliation{%
 Department of Physics, Dibrugarh University, Dibrugarh \\
 Assam, India, 786004}

\author{Tonmoyee Gogoi \orcidlink{0009-0009-9195-3627}}%
 \email{tonmoyeegogoi2062@gmail.com}
\affiliation{%
 Department of Physics, Dibrugarh University, Dibrugarh \\
 Assam, India, 786004}%

\author{Kakoli Bhuyan \orcidlink{0009-0007-6949-890X}}%
 \email{bhuyankakoli04@gmail.com}
\affiliation{%
 Department of Physics, Dibrugarh University, Dibrugarh \\
 Assam, India, 786004}%

\author{Prabwal Phukon \orcidlink{0000-0002-4465-7974}}%
 \email{prabwal@gmail.com}
\affiliation{%
 Department of Physics, Dibrugarh University, Dibrugarh \\
 Assam, India, 786004}%
 \affiliation{Theoretical Physics Divison, Centre for Atmospheric Studies, Dibrugarh University, Dibrugarh, Assam, India 786004}

\begin{abstract}
The Frolov black hole (BH) is a charged extension of the Hayward BH, having regularity at the central point $r = 0$ and an asymptotically Schwarzschild form for large values of $r$. Such a BH is parameterized by a length scale parameter, \( \alpha_0 \). In this paper, we analyze the thermodynamic properties, null and timelike geodesics, and shadows of a Frolov BH immersed in a quintessence field. Our results indicate that the smaller BH is locally thermodynamically stable yet globally unstable at all horizon radii. Neither the quintessence parameter nor the other model parameters like the charge $q$ and length scale parameter $\alpha_0$ change this global instability. We extend the study of the null and timelike geodesics to the vicinity of the BH by analyzing how the geodesic motion depends on the model parameters. Finally, we analyze the shadow of the BH system and find that the shadow radii are sensitively dependent on model parameters. In contrast, the influence of the quintessence parameter itself on the size of the shadow is found to be rather weak.
\end{abstract}

\maketitle
%\newpage
%\tableofcontents
\section{Introduction}
\label{intro}
Black holes (BHs) are interesting theoretical and astrophysical objects that have become an intriguing topic of discussion over the years. The importance of BHs accelerated in recent years after the first direct observation of the BH shadow by the Event Horizon Telescope (EHT) collaboration in the center of the M87 galaxy \cite{EventHorizonTelescopeCollaboration2022May}. Historically speaking, the first known solutions of the Einstein field equations, given by Schwarzschild, give rise to the possibility of a spacetime that embeds a spherically symmetric, chargeless and non-rotating body, known as the Schwarzschild BH later on. Another generalization of the uncharged Schwarzschild BH to a charged one is the Riessner-Nordstr\"{o}m BH. Some of other types of BHs are the Kerr BH, which represents the rotating and uncharged BH solution, Kerr-Newman BH - which is a charged and rotating axisymmetric BH solution. BHs exhibit several interesting physical properties, like thermodynamics and optical properties like shadows, lensing, etc.

Black hole thermodynamics (BHT) \cite{Page2005Sep,Bardeen1973Jun} is a growing field of inquiry in BH physics. Hawking radiation, also known as thermal fluctuations \cite{Hawking1974Mar,Hawking1971May,Page2005Sep,Wald2001Jul,Bardeen1973Jun,Hawking1975Aug}, is a unique phenomenon associated with black holes. Hawking's theory of BH radiation asserts that BHs are not absolutely black, but instead produce radiation at a temperature inversely proportional to their mass. The radiation originates via quantum mechanical processes near the event horizon, where virtual particles are continually created and destroyed \cite{Hawking1975Aug}.
These particles can sometimes leave the BH as radiation that causes the BH to gradually lose mass. This is known as BH evaporation. This process takes place through a relativistic quantum mechanical mechanism and may be crucial in offering promising insights towards a quantum theory of gravity. As per the laws of BHT, BH systems can be treated as thermodynamic systems that follow the usual laws of thermodynamics. In the same manner, a BH can also possess analogous thermodynamic properties like temperature, entropy, free energy, etc. Recent studies on this topic have been carried out in different possible backgrounds. For instance, Ditta et al. \cite{Ditta2023Jun} studied the thermodynamics of
BHs that resemble charged tori. Acoustic Schwarzschild BHs in the framework of extended thermodynamic phase space, those admitting phase transition, were studied by Yasir et al. \cite{Yasir2024Jan}. Mahapatra and Banerjee \cite{Mahapatra2023Feb} addressed the thermodynamics of rotating hairy black holes through gravitational coupling. In the context of Maxwell electrodynamics, Hendi et al. \cite{Hendi2021Dec} examined the thermodynamics and phase transitions of a four-dimensional rotating Kaluza–Klein black hole. In a higher-derivative theory, Singh et al. \cite{Singh2022Mar} developed a black hole solution in Lee-Wick gravity with a point source and then studied the thermodynamics of such a BH system. Simovic and Soranidis \cite{Simovic2024Feb} used both Euclidean path integral and Hamiltonian approaches to investigate the thermodynamic parameters of the Hayward regular BH in asymptotically anti-de Sitter, Minkowski, and de Sitter spacetimes. Many other important works can be found in Refs. \cite{Cassani2024May,Sadeghi2024Jan,Aref'eva2024Feb,Abbas2024Apr,
Ladghami2024May,Sokoliuk2024Jan,Yang2024Apr,Paul2024Apr,Ladghami2024May1,Ali2024Apr,
Rakic2024Jun,Sadeghi2024Jan1,Davies2024Apr,Kruglov2024Feb,Ballesteros2024Feb,
Yue2024Jul,Capozziello2024Apr,Battista2024Jan,Battista2022Dec}.

The theory of GR is known to be UV incomplete in both classical and quantum backgrounds; in other words, the theory is not free from singularities. The well known BH sytems, like Schwarzschild, Riessner Nordstr\"{o}m and Kerr, exhibit curvature singularities at the centre. This offered the need for a modification in Einstein's GR if one wants a UV complete theory. There have been many attempts to formulate alternative versions of a UV complete theory, but these modifications have their own problems; for instance, introducing higher-order curvature and derivative terms into the gravitational action leads to the so-called ghost instabilities, where ghosts typically mean unphysical degrees of freedom. In a notable work, V.L. Frolov in 2016 \cite{Frolov2016Nov}, worked out the possibility of non-singular BH solutions without modifying the theory of gravity by constructing metrics through several intuitive assumptions. Frolov, in his original paper, put forward several possibilities, including the modified Hayward solution by generalising to the charged case. He assumed that there should exist a critical energy scale $\mu$, which in turn relates to the length scale parameter $\alpha_0$ (In his original paper, he assumed it to be $l$) by $ \alpha_0 = \mu^{-1}$. Therefore, in his construction, apart from the mass of the BH, there exists an additional parameter $\alpha_0$, which essentially determines the scale where the modification of Einstein's equations becomes important. In a more technical sense, at a scale where the curvature scalar $R$ becomes comparable to $\alpha_0^{-2}$. The second assumption is that one can utilize the usual metric tensor $g_{\mu \nu}$ and also that there exists a length scale $\lambda$ at which quantum gravity effects become significant, whose magnitude lies way below the parameter $\alpha_0$. These assumptions render the solution non-singular. Following this work, \cite{Song2024Aug} investigated quasinormal modes under scalar perturbations, which were shown to be stable with time decay behaviour, influenced by quantum gravity effects. Also, Kumar et al investigated the shadow and lensing properties of a Frolov BH \cite{Kumar2019Dec} by constraining the BH parameters for the M87* shadow observational data.

In this paper, we shall discuss the thermodynamic properties, geodesics and shadows of a Frolov BH surrounded by a quintessence field. The paper is organized as follows: In Section \ref{sec2}, we discuss the theoretical framework of Frolov BH in the presence of quintessence. In Section \ref{sec3}, we discuss the thermodynamic properties and the stability of the Frolov-quintessence BH system. In Section \ref{sec4}, we derive the effective potential for both null and timelike particles of the Frolov-quintessence BH system and hence study the null and timelike geodesics. In Section \ref{sec5}, we obtain the shadow cast by the Frolov-quintessence BH. Finally, in Section \ref{conc} we summarize the results of our study.

\section{Frolov Black Hole}
\label{sec2}
The charged generalization to a Hayward BH is known as the Frolov BH. The metric associated with the Frolov BH is given by \cite{Song2024Aug}
\begin{equation}
ds^2 = - f(r) dt^2 + f(r)^{-1} dr^2 + r^2 d\Omega^2,
\label{line_el}
\end{equation}
where 
\begin{equation}
f(r) = 1 - \frac{(2Mr - q^2)r^2}{r^4 + (2Mr + q^2)\alpha_0^2},
\label{lapse}
\end{equation} and $d\Omega^2 = d\theta^2 + \sin^2 \theta d\phi^2$, 
with $M$ being the ADM mass of the BH. The Frolov BH is characterized by the cosmological constant $\Lambda = 3/\alpha_0^2$, where $\alpha_0$ is the Hubble length (length scale parameter) \cite{Song2024Aug,Hayward2006Jan}. The Hubble length manifests itself as a Universal hair and is constrained by $\alpha_0 \le \sqrt{16/27}M$. In the following analysis, we shall set $M = 1$ for the ease of calculation. The charge parameter $q$ satisfies the constraint $0\le q \le 1.$ It is easy to find that in the limiting value of $q \to 0$, the Frolov BH reduces to the Hayward solution. As $\alpha_0 \to 0$, the trivial case of Reissner-N\"{o}rdstrom BH solution is obtained. Further, as both $q$, $\alpha_0 \to 0$, the Schwarzschild solution can be recovered. On the other hand, Kiselev suggested that a quintessence field can be incorporated in the framework of a BH \cite{Kiselev2003Mar}. The quintessence field should satisfy the relations
\begin{equation}
T^{\phi}_{\phi} = T^{\theta}_{\theta} = -\frac{1}{2}(3w + 1) T^{r}_{r} = \frac{1}{2}(3w + 1) T^{t}_{t}
\end{equation}
where $w$ represents the equation of state (EoS) parameter of the quintessence field. Typically, the EoS parameter $w$ for quintessence lies within the range $-1 < w < -1/3$.

Inspired by this, in this work, we shall work with a combined framework of a BH system and study its various thermodynamic and optical properties from different perspectives in the presence of quintessence. Following the idea of Kiselev, the effective metric can be constructed incorporating the effect of a quintessence field by simply adding the term $-c/r^{3w+1}$ into the BH metric, Eq. \eqref{lapse}, to get 
\begin{equation}
f(r) = 1 - \frac{(2Mr - q^2)r^2}{r^4 + (2Mr + q^2)\alpha_0^2} - \frac{c}{r^{3w + 1}},
\label{lapse_quin}
\end{equation}
in the modified lapse function given by Eq. \eqref{lapse_quin}. Some of the recent related works carried out in this setting are \cite{Zeng2020Nov,Mustafa2022Dec,Belhaj2020Oct,Chen2022Apr,Toshmatov2017Feb} (and references therein).  In our subsequent work, we shall choose the case where we fix $\omega = -2/3$. Also, $c$ is a constant parameter that effectively describes a coupling of the quintessence matter with the BH system. In an overall sense, our model Eq. \eqref{lapse_quin} has altogether three parameters of dependence, i.e. the quintessence coupling parameter $c$, the Hubble length scale $\alpha_0$, and charge $q$. As mentioned before, the existence ranges of $\alpha_0$ and $q$ are already predefined. This leaves us with the quintessence parameter $c$ as an unconstrained free parameter. In the subsequent analysis, we shall try to determine the allowed values of $c$ concerning different physical aspects.
%\begin{figure*}
%\centerline{\includegraphics[scale=0.5]{fr_plot.pdf}}
%\caption{The lapse function is plotted for different combination of the parameter quintessence parameter $c$, $\alpha$, and charge $q$. }
%\label{fr_plot}
%\end{figure*}
\begin{figure*}
\centerline{\includegraphics[scale=0.48]{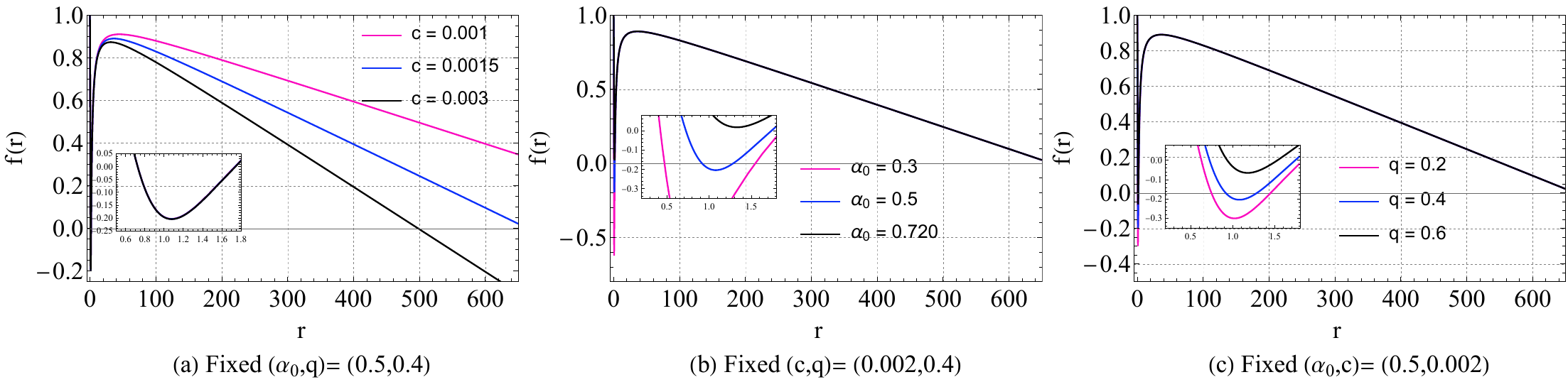}}
\caption{The lapse function is plotted for different combinations of the parameters $c$, $\alpha_0$, and charge $q$. }
\label{fr_plot1}
\end{figure*}
As a preliminary check, we start by plotting Eq. \eqref{lapse_quin} for different values of $c$, $\alpha_0$ and $q$ in Fig. \ref{fr_plot1}. From the plot, we observe that the BH system exhibits three horizons $r_h^{(1)}$, $r_h^{(2)}$ and $r_h^{(3)}$, which represent smaller, intermediate and large BHs, respectively. It is interesting to notice that the parameters $q$ and $\alpha_0$ seem to influence the occurrence of horizons at smaller and intermediate radii, whereas a feeble effect is seen at larger radii. In contrast, the quintessence parameter $c$, tends to impact feebly at smaller and intermediate horizon radii, but affects drastically in the occurrence of large horizon radii (see Fig. \ref{fr_plot1}). It can also be seen that corresponding to the value of $\alpha_0 = 0.720$, there exists a single horizon, known as the extremal horizon. Out of the three horizons, two of them are fairly close to each other, while the third lies at extremely large distances (which can be termed as the cosmological horizon). The far away horizon may be real, but is likely to be irrelevant in the context of local BH thermodynamic phenomenon. Moreover, the existence of the innermost horizon may or may not be thermodynamically relevant, which will be investigated in detail in the next section. 

\section{Thermodynamical Parameters}
\label{sec3}
In this section, we shall obtain the thermodynamic functions relevant to the Frolov black hole system in the presence of quintessence matter. Let us begin by obtaining the Hawking temperature $T_h$ through the direct relation with the metric function  Eq. \eqref{fr_plot1}, which is 
\begin{equation}
\begin{aligned}
T &= \frac{1}{4 \pi  r_h^3 \left(r_h^4+2 \alpha _0^2 q^2\right)} \left[ q^2 \left(\alpha _0^4+\alpha _0^2 \left(\alpha _0^2 c^2-4\right) r_h^2 \right. \right. \\& \left. \left. \hspace{1cm}+2 \alpha _0^2 c r_h^3-2 \alpha _0^4 c r_h-r_h^4\right)+r_h^4 \left(-3 \alpha _0^2 \right. \right. \\& \left. \left. \hspace{1cm}+\left(1-3 \alpha _0^2 c^2\right) r_h^2+6 \alpha _0^2 c r_h-2 c r_h^3\right)\right],
\end{aligned}
\label{temp}
\end{equation}
where $r_h$ is the horizon radius.
\begin{figure*}[htb]
\centerline{\includegraphics[scale=0.55]{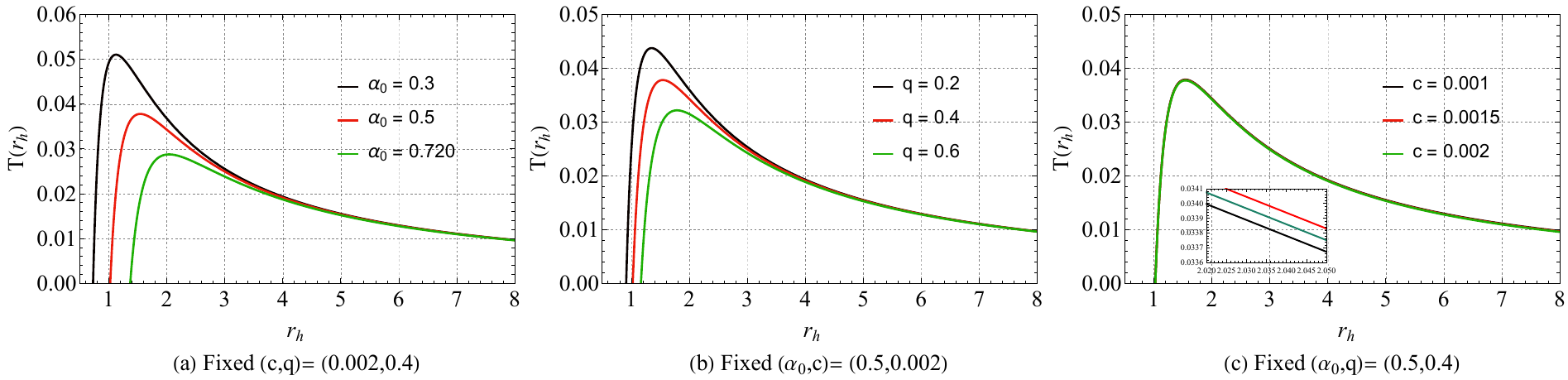}}
\caption{The Hawking temperature is plotted for different values of the model parameters $\alpha_0$, $q$ and the quintessence parameter $c$.}
\label{temp_plot}
\end{figure*}
Fig. \ref{temp_plot} illustrates the behaviour of the Hawking temperature $T$ as a function of the horizon radius \( r_h \), the model parameters $\alpha_0$, $q$ and the quintessence parameter $c$. The plot illustrates that the temperature initially increases with increasing horizon radius, reaches a maximum value, and then decreases as the horizon radius increases further. Additionally, it is seen that the peak value of the Hawking temperature decreases with increasing values of $\alpha_0$, while keeping $c$ and $q$ fixed. Also, the decrease in the peak value of temperature occurs with increasing values of charge. Interestingly, the quintessence parameter does not seem to have a drastic impact on the temperature variation and the peak values.  Therefore, it is quite safe to state the BH with a lower charge for fixed values of $\alpha_0$ and $c$, or a lower value of $\alpha_0$ with a fixed value of $q$ and $c$ tends to exhibit higher temperature. Moreover, the horizon radius $r_h$ at which the temperature peak occurs shifts to larger values as the value of $\alpha_0$ or $q$ increases. Also, as the horizon radius increases further, the temperature asymptotically vanishes. At radii below some minimal horizon radius (corresponding to $T = 0$), the temperature takes on negative values, referring to no physical BH states.

Again, setting $f(r_h) = 0$ leads us to obtain the Arnowitt-Deser-Misner (ADM) mass ($M$) of the black hole given by
\begin{equation}
M = \frac{-\alpha _0^2 c q^2 r_h-c r_h^5+q^2 r_h^2+r_h^4+\alpha _0^2 q^2}{2 r_h \left(-\alpha _0^2+\alpha _0^2 c r_h+r_h^2\right)},
\label{mass}
\end{equation}
Clearly, in the limiting values of $\alpha_0, c, q \, \to 0$, the mass of Schwarszchild BH is recovered,
\begin{equation}
M_{sch} = \frac{r_h}{2},
\label{sch_mass} 
\end{equation}
\begin{figure*}
\centerline{\includegraphics[scale=0.45]{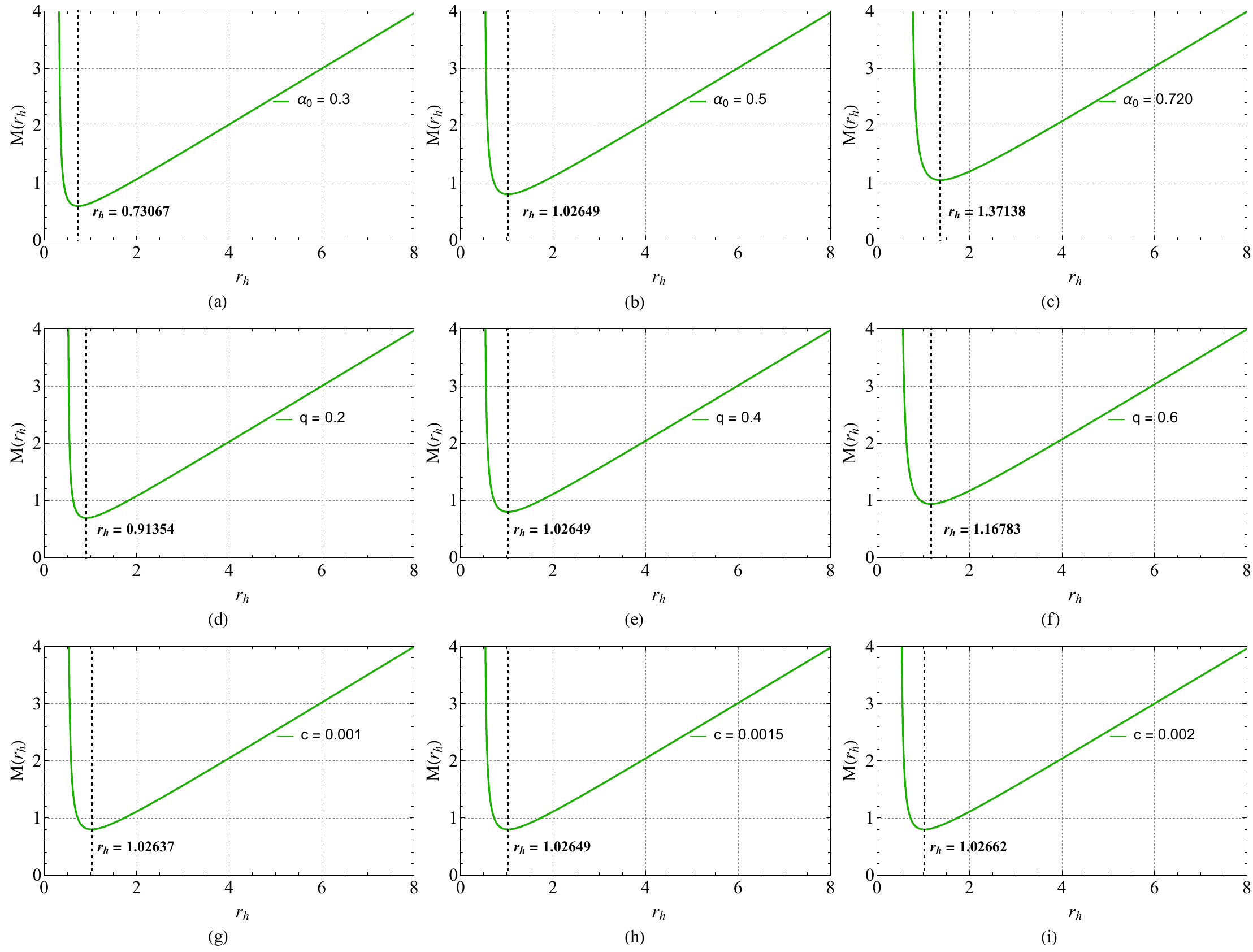}}
\caption{The ADM mass is plotted for different combinations of the parameters $c$, $\alpha_0$, and charge $q$.}
\label{mass_plot}
\end{figure*}
From Fig. \ref{mass_plot}, the vertical lines in the plots indicate the horizon radius at which the temperature goes to zero. Therefore, the BHs with horizon radii less than the radii that correspond to $T=0$, (let us call this point $r_{T=0}$) can be ruled out as they would refer to BH states with negative temperature. Beyond $r_{T=0}$, we see that the ADM mass increases monotonically with increasing horizon radius. 
The entropy of the BH can be calculated as
\begin{equation}
S_{BH} = \int \frac{dM}{T} = \pi r_h^2,
\label{bh_entropy}
\end{equation}

Analysis of certain parameters about thermodynamic stability in the study of BH systems is crucial, as it reveals both local and global stability characteristics of the system. For local stability, a key thermodynamic property is the specific heat capacity ($C$) as it responds to how a BH reacts to small perturbations in temperature. A positive value of specific heat is an indication that the system is locally stable and can absorb heat with a moderate temperature, while negative values of specific heat indicate local instability in the form of increasing temperature with the decrease of mass.

Global stability requires the Helmholtz free energy to be in the negative domain. It would further determine if the black hole configuration is thermodynamically preferred over other configurations. At a given temperature, Helmholtz free energy that is smaller implies global stability in the system for being a minimum energy level at constant temperature. Analysis of such parameters gives an overall view of the thermodynamic stability of black hole systems at both the local and global levels. 

The specific heat can be obtained as
\begin{widetext}
\begin{equation}
C = \frac{\partial M}{\partial T} = -\frac{2 \pi  r_h^2 \left(r_h^2-\alpha _0^2\right) \left(q^2 \left(\alpha _0^4-4 \alpha _0^2 r_h^2-r_h^4\right)-3 \alpha _0^2 r_h^4+r_h^6\right)}{-q^2 \left(3 \alpha _0^6+21 \alpha _0^2 r_h^4-11 \alpha _0^4 r_h^2+3 r_h^6\right)-6 \alpha _0^2 r_h^6-3 \alpha _0^4 r_h^4+r_h^8},
\label{spheat}
\end{equation}
\end{widetext}
\begin{figure*}
\centerline{\includegraphics[scale=0.45]{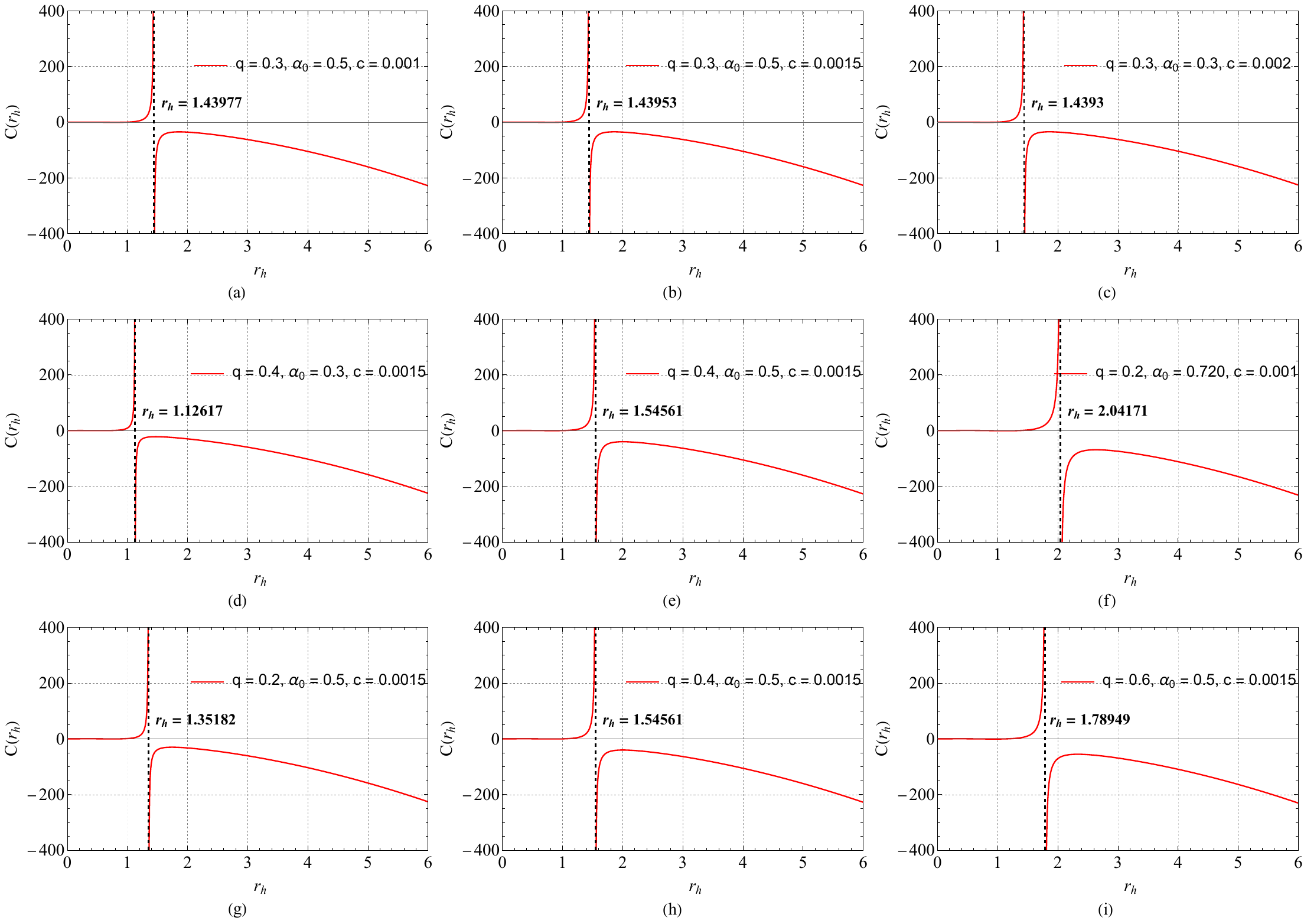}}
\caption{The specific heat is plotted for different combinations of $\alpha_0$, $q$ and $c$.}
\label{spheat_plot}
\end{figure*}
By plotting the specific heat $C$ with the horizon radius $r_h$ as shown in Fig. \ref{spheat_plot}, we see the dependence of the position of Davies point on the parameters $\alpha_0$, $q$ and $c$.  Concerning the quintessence parameter $c$, we observe that for lower values of $c$, the Davies point occurs at higher values of the critical horizon radius $r_h^c$. This phase transition makes the BH shift from a locally stable BH state to an unstable BH state. Therefore, at a horizon radius $r_{T=0} < r_h < r_h^c$, the black hole is locally stable, whereas it is unstable for $r_h > r_h^c$.  In contrast, from the context of dependence on $\alpha_0$, the Davies point shifts toward a larger horizon radius with increasing values of $\alpha_0$. A similar trend is seen as the charge $q$ increases. 

Next, to examine the global stability of our BH system, we analyze the Helmholtz free energy ($F$) in our model. A positive value of $F$ indicates the global instability of the thermodynamic system, whereas a negative value indicates global stability. The Helmholtz free energy at the horizon can be calculated as
\begin{equation}
\begin{aligned}
F &= M - TS \\&= \frac{1}{4 r \left(-\alpha _0^2+\alpha _0^2 c r+r^2\right)^2} \left[q^2 \left(-3 \alpha _0^4+r^2 \left(4 \alpha _0^2-3 \alpha _0^4 c^2\right) \right. \right. \\& \left. \left. -2 \alpha _0^2 c r^3+6 \alpha _0^4 c r+3 r^4\right) +r^4 \left(\alpha _0^2+r^2 \left(\alpha _0^2 c^2+1\right)\right. \right. \\& \left. \left. \hspace{6cm}-2 \alpha _0^2 c r\right)\right]
\label{free_energy}
\end{aligned}
\end{equation}
\begin{figure*}
\centerline{\includegraphics[scale=0.45]{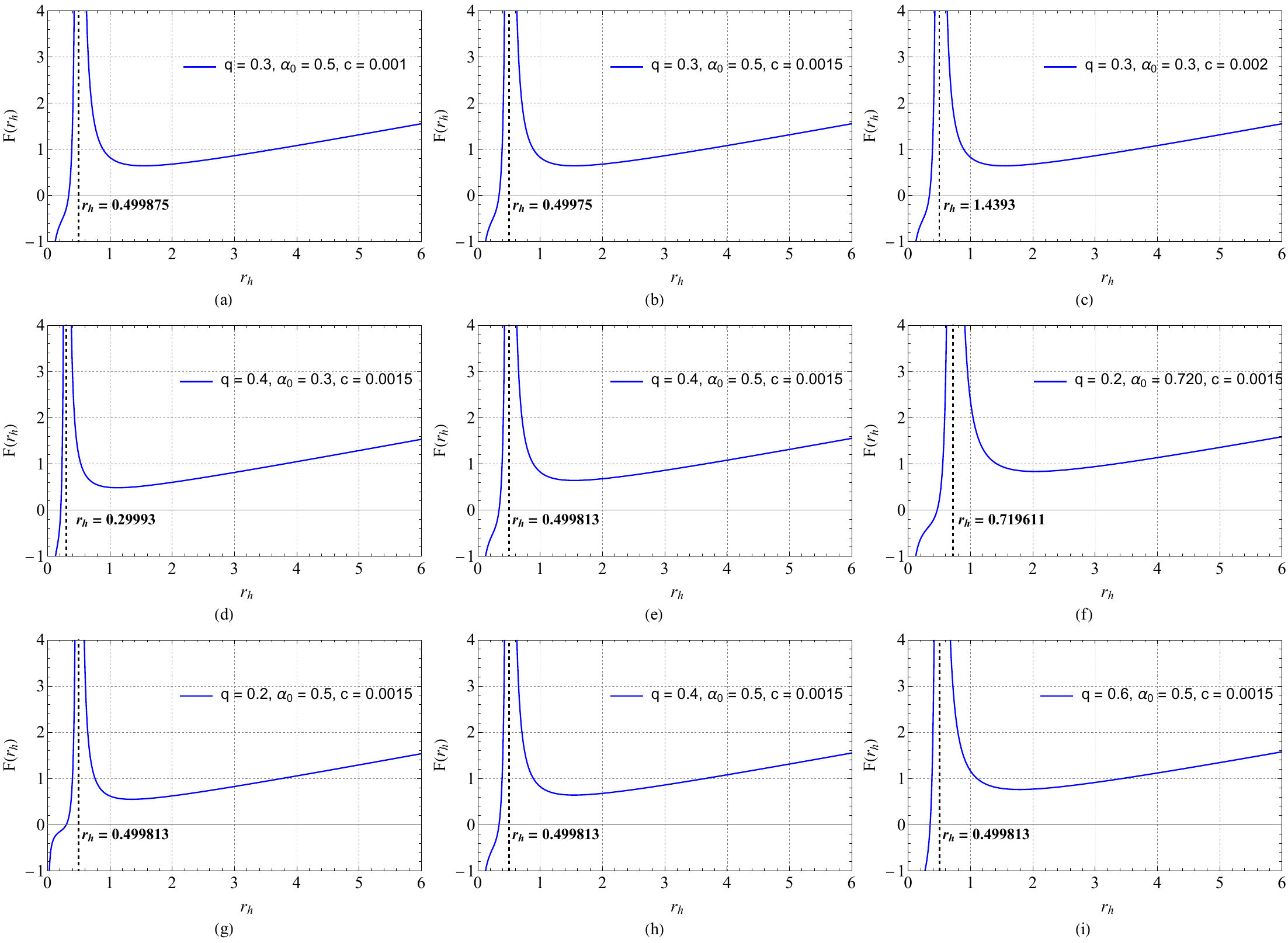}}
\caption{The free energy is plotted for different values of $\alpha_0$, $q$ and $c$.}
\label{free_energy_plot}
\end{figure*}
Fig. \ref{free_energy_plot} shows the evolution of the Helmholtz free energy for our system for different combinations of the model parameters and quintessence parameter. The vertical lines in the plots mark the Hawking-Page transition points where the phase transition takes place. As it was mentioned earlier, we have set the lower limit of the allowed horizon radius for the positivity of the temperature of the black hole states. The free energy in our system takes positive values for all allowed BH states, which indicates that the BH system in our model is globally unstable. 
\section{Effective Potential: Null and Timelike Geodesics}
\label{sec4}
In spherically symmetric spacetimes, the functions $A(r)$ and $B(r)$ are a redshift factor and a spatial curvature that determines the motion of a particle (massless or massive) within the gravitational field of a BH. In such a spacetime, two Killing vectors intrinsic to the metric could be found, which are associated closely with time translation and rotational symmetry, yielding two conserved quantities respectively along the geodesics. Specifically, the time-translational Killing vector produces a conserved energy, $E = A(r) \dot{t}$, that remains constant along the geodesic of the particle. Similarly, the Killing vector associated with rotational symmetry represents a conserved angular momentum, which makes the motion constrained to orbits (parabolic, hyperbolic or elliptical, depending on their eccentricities). These intrinsic symmetries lead to conserved quantities that play a crucial role in determining the equations of motion. Further, one can derive the effective potential controlling both null and timelike particle trajectories and see whether the orbits are stable or unstable or whether precession of timelike orbits occurs.

To obtain the geodesic equations of motion, let us start from the generic form of a spherically symmetric spacetime metric given by
\begin{equation}
ds^2 = - A(r)^2 dt^2 + B(r)^2 dr^2 + r^2 d\theta^2 + r^2 \sin^2 \theta d\phi^2,
\label{gen_met} 
\end{equation}
The metric given by \eqref{gen_met} possesses time translational and rotational symmetry, which implies that the Killing vector associated with it produces conserved quantities along the geodesics, given by 
\begin{equation}
K_\mu \dot{x}^\mu = \text{ constant}.
\label{kill}
\end{equation}
Here, the `$.$' represents the differentiation with respect to the affine parameter $\lambda$. The Killing vectors corresponding to the time translational symmetry and rotational symmetry are
\begin{equation}
K_\mu = ( - A(r), 0, 0, 0),
\label{time_killing}
\end{equation}
and 
\begin{equation}
K_\mu = (0, 0, 0, r^2 \sin^2 \theta),
\label{sph_killing}
\end{equation} respectively.
From the Eqs. \eqref{gen_met} and \eqref{time_killing}, we can obtain the E-Equation as
\begin{equation}
E = A(r)\dot{t} = \text{constant}.
\label{E-eqn}
\end{equation}
Keeping notice of the fact that we are dealing with rotational symmetry, we have the freedom to restrict the observer's point of view to the equatorial plane by fixing $\theta = \pi/2$. This leads us to obtain the L-Equation from Eqs. \eqref{gen_met} and \eqref{sph_killing} as
\begin{equation}
L = r^2 \dot{\phi} = \text{constant},
\label{L-eqn}
\end{equation}
It can also be seen that the norm of the tangent vector to the geodesic is also a conserved quantity, which means
\begin{equation}
\epsilon = - g_{\mu \nu} \dot{x}^\nu \dot{x}^\mu,
\label{eps}
\end{equation}
where $\epsilon = 0, 1$ for null and timelike geodesics respectively. Using the metric \eqref{gen_met} in Eq. \eqref{eps} we get
\begin{equation}
-\epsilon = -A(r) \dot{t}^2 + B(r) \dot{r}^2 + r^2 \dot{\phi}^2.
\label{eps1}
\end{equation}
Eq. \eqref{eps1} can also be written in the form
\begin{equation}
\dot{r}^2 = \frac{E^2}{A(r) B(r)} - \frac{L^2}{r^2 B(r)} - \frac{\epsilon}{B(r)}.
\label{rdotsq}
\end{equation}
Now in order to derive the geodesic equations, we may utilize the Lagrangian given by 
\begin{equation}
\mathcal{L} = \frac{1}{2}g_{\mu \nu} \dot{x}^\alpha \dot{x}^\beta = \frac{1}{2}\left(-A(r) \dot{t}^2 + B(r) \dot{r}^2 + r^2 \dot{\phi}^2\right).
\label{Lag}
\end{equation}
By using the Euler-Lagrange Equation in the $r$-coordinate
\begin{equation}
\frac{d}{d\lambda}\left( \frac{\partial \mathcal{L}}{\partial \dot{r}}\right) = \frac{\partial \mathcal{L}}{\partial r},
\label{Lagr}
\end{equation}
we obtain
\begin{equation}
\dot{p}_r = \frac{1}{2}\left(- \frac{\partial A(r)}{\partial r}\dot{t}^2 + \frac{\partial B(r)}{\partial r}\dot{r}^2 + 2r \dot{\phi}^2 \right),
\label{prdot}
\end{equation}
using the conjugate momentum in $r$-coordinate 
\begin{equation}
p_r = \frac{\partial \mathcal{L}}{\partial \dot{r}}= \dot{r}B(r)
\label{pr}
\end{equation}

\subsection{Null Geodesics}
By using Eqs. \eqref{E-eqn}, \eqref{L-eqn}, \eqref{prdot} and \eqref{pr} we can obtain the set of equations of motion for null-geodesics in the general spherically symmetric spacetime \eqref{gen_met} as:
\begin{equation}
\begin{aligned}
\dot{t} &= E A(r)^{-1} \\
\dot{\phi} &= \frac{L}{r^2} \\
\dot{r} &= p_r B(r)^{-1} \\
\dot{p}_r &= \frac{1}{2}\left(- \frac{E^2}{A(r)^2} \frac{\partial A(r)}{\partial r} + \frac{p_r^2}{B(r)^2} \frac{\partial B(r)}{\partial r} + \frac{2L^2}{r^3}\right).
\end{aligned}
\label{geod_eqs}
\end{equation}
From Eq. \eqref{rdotsq}, we may find
\begin{equation}
\frac{1}{2}\dot{r}^2 + V_{eff} = \frac{E^2}{2},
\label{eff1}
\end{equation}
where \begin{equation}
V_{eff} =  \frac{L^2}{2r^2} B(r)^{-1} + \frac{\epsilon}{B(r)}.
\label{eff2}
\end{equation}
Utilizing the metric \eqref{lapse}, after comparing with the metric \eqref{gen_met}, we get $A(r) = f(r)$, $B(r) = f(r)^{-1}$. Therefore, the system of Eqs. \eqref{geod_eqs} takes the form

\begin{widetext}
\begin{equation}
\begin{aligned}
\dot{t} &= E \left(1-c r-\frac{r^2 \left(2 M r-q^2\right)}{\alpha _0^2 \left(2 M r+q^2\right)+r^4} \right)^{-1} \\
\dot{\phi} &= \frac{L}{r^2} \\
\dot{r} &= p_r \left(1 -c r-\frac{r^2 \left(2 M r-q^2\right)}{\alpha _0^2 \left(2 M r+q^2\right)+r^4}\right) \\
\dot{p}_r &= \frac{\left(r^5 \left(c r^3-2 M r+2 q^2\right) + \alpha_0^4 c \left(2 M r + q^2\right)^2 
+ 2 \alpha_0^2 r \left(q^2 \left(c r^3 + 2 M r\right) 
+ 2 M r^2 \left(c r^2 + 2 M\right) - q^4\right)\right)}{2 \left(\alpha_0^2 \left(2 M r + q^2\right) + r^4\right)^2} \\
&\times \frac{\left(p_r^2 \left(r^2 \left(-c r^3 - 2 M r + q^2 + r^2\right) 
- \alpha_0^2 (c r - 1) \left(2 M r + q^2\right)\right)^2 
+ e^2 \left(\alpha_0^2 \left(2 M r + q^2\right) + r^4\right)^2\right)}{\left(r^2 \left(r \left(-c r^2 - 2 M + r\right) 
+ q^2\right) - \alpha_0^2 (c r - 1) \left(2 M r + q^2\right)\right)^2} \\
&+ \frac{L^2}{r^3}.
\end{aligned}
\label{geod_eqs_mod}
\end{equation}
\end{widetext}

\begin{figure*}[tbh]
\centerline{\includegraphics[scale=0.3]{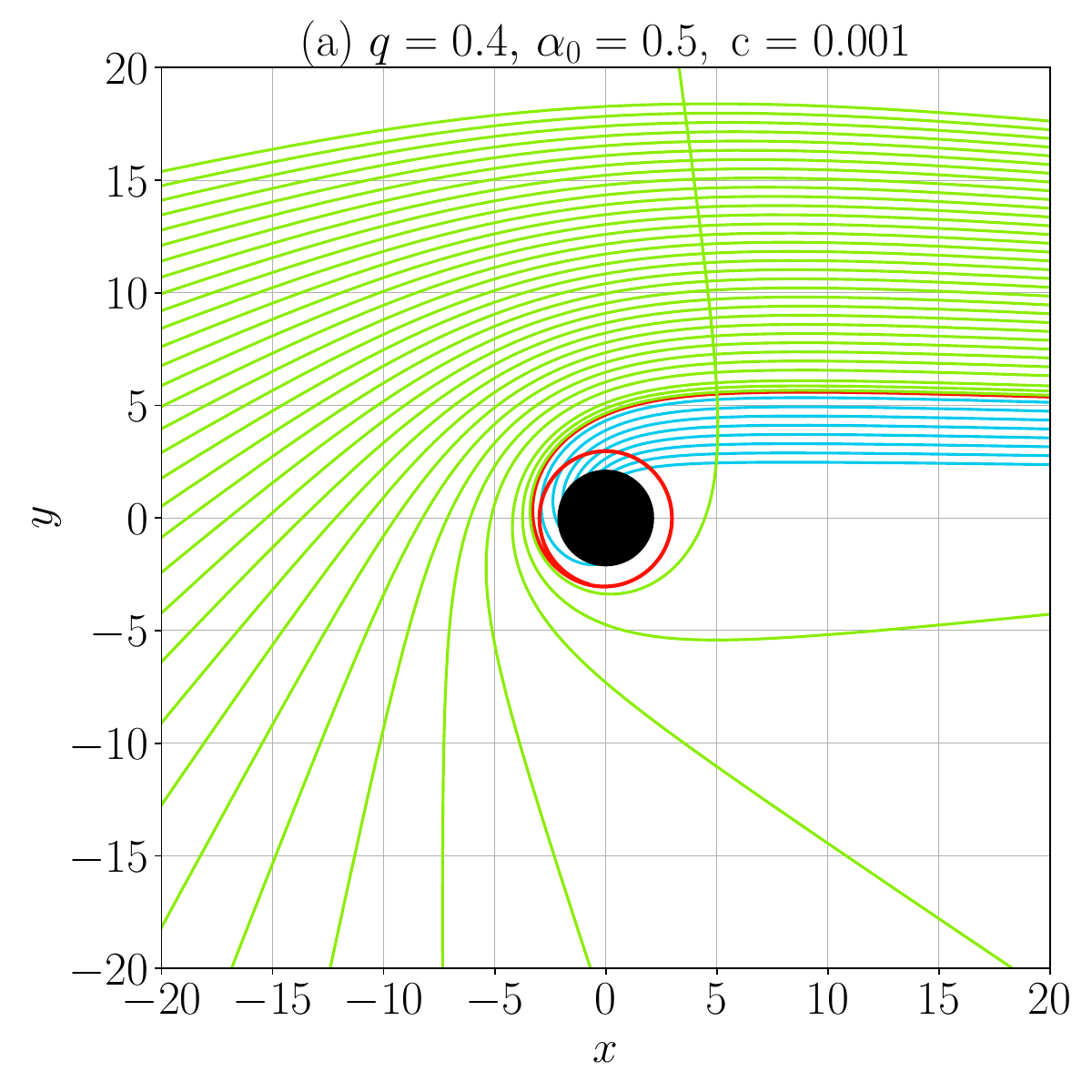}\hspace{-0.1cm}\includegraphics[scale=0.3]{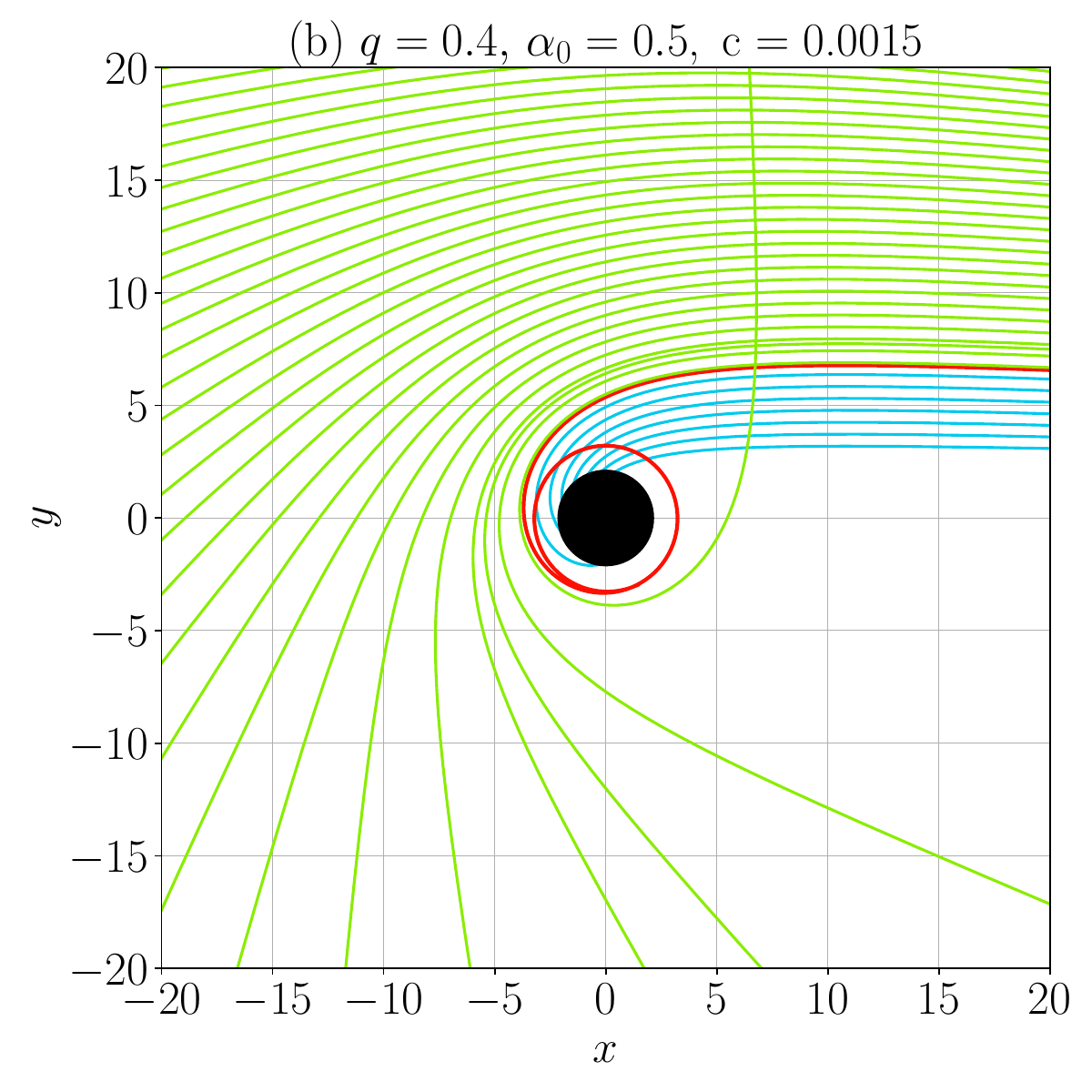} \hspace{-0.1cm}\includegraphics[scale=0.3]{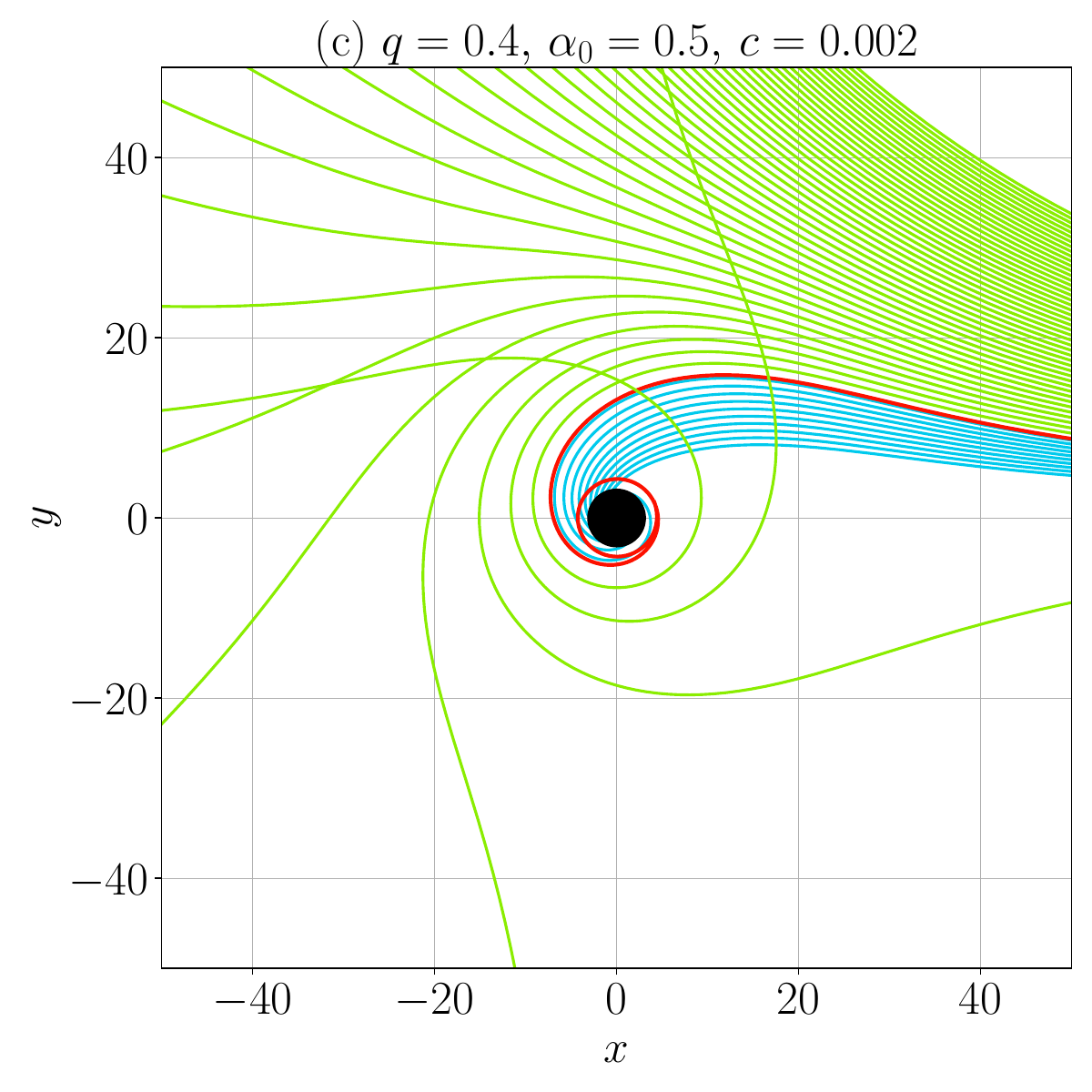}}
\centerline{\includegraphics[scale=0.3]{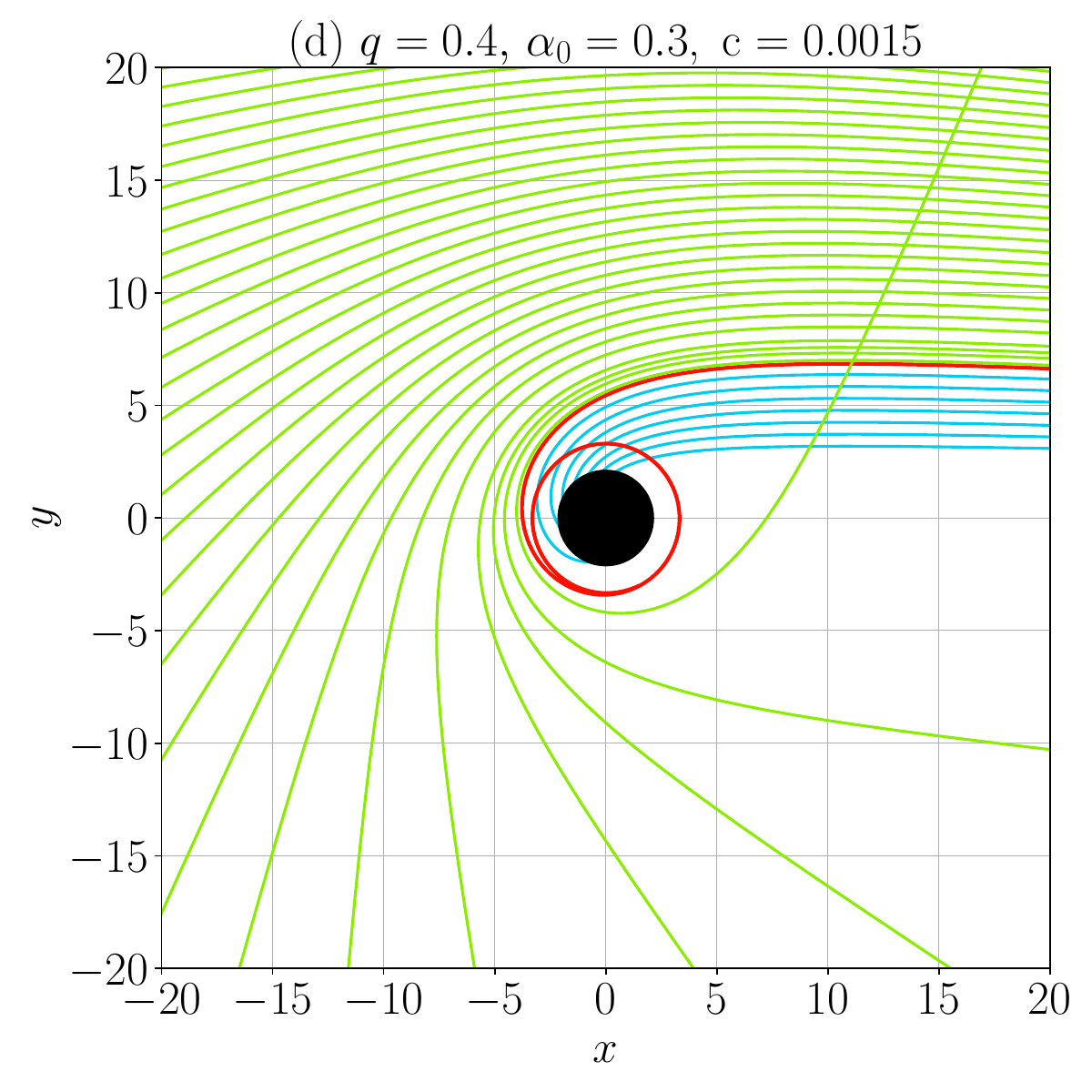}\hspace{-0.1cm}\includegraphics[scale=0.3]{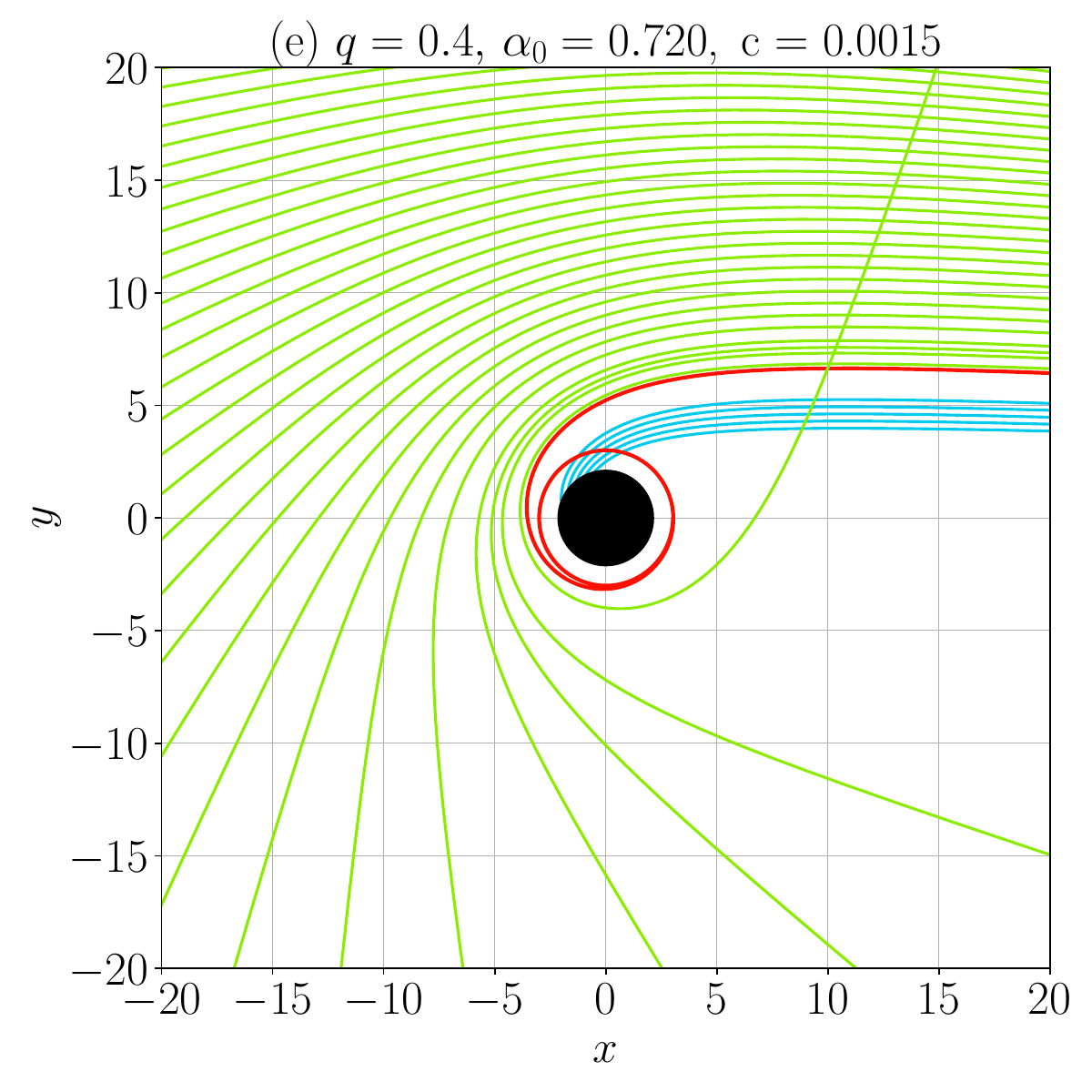} \hspace{-0.1cm}\includegraphics[scale=0.3]{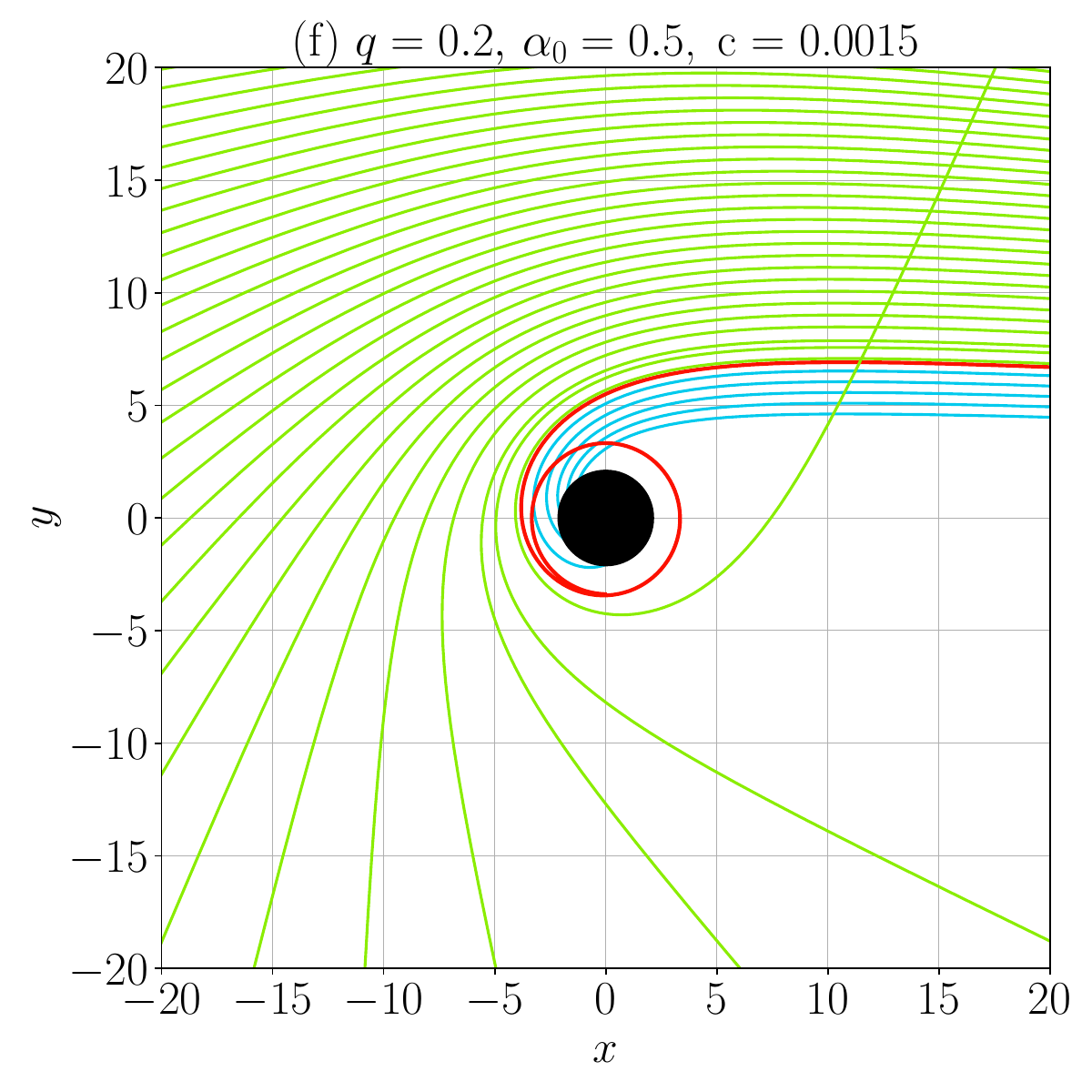}}
\centerline{\includegraphics[scale=0.3]{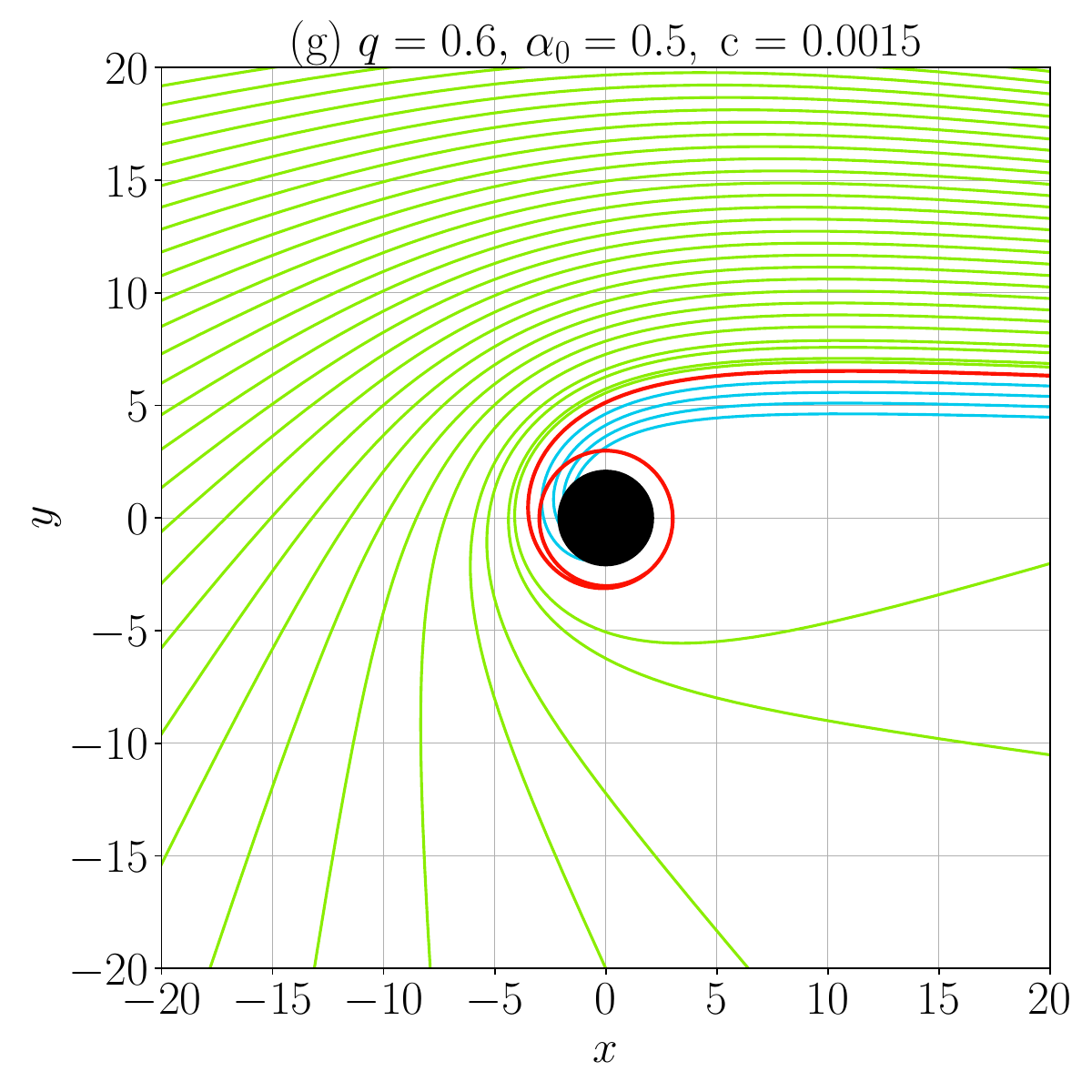}}
\caption{Null geodesics for different combinations of the model parameters}
\label{null_geod}
\end{figure*}

\begin{figure*}[tbh]
\centerline{\includegraphics[scale=0.48]{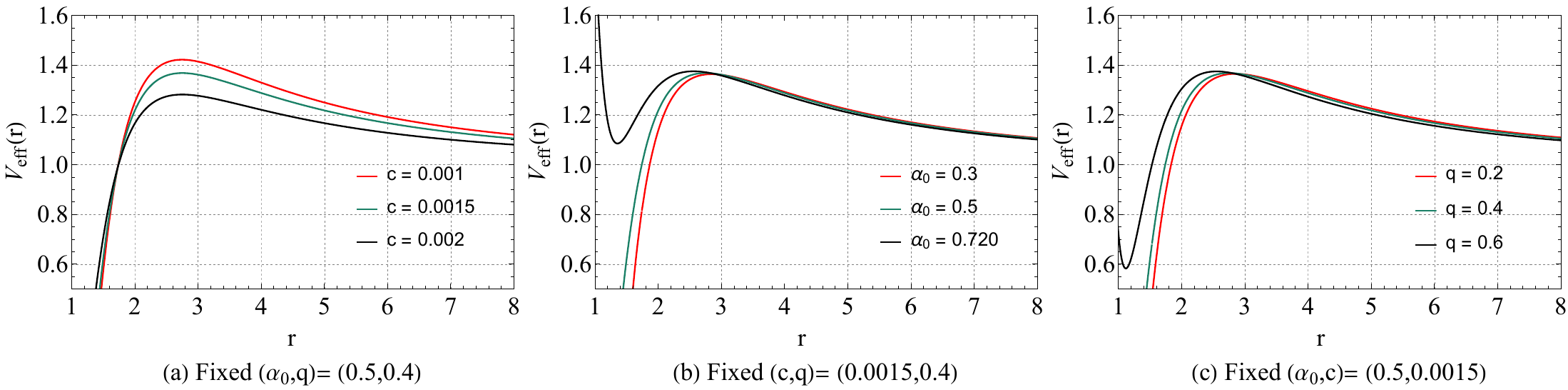}}
\caption{The plot of effective potential is shown for different combinations of the parameters $c$, $\alpha_0$, and charge $q$.}
\label{eff_pot_plot}
\end{figure*}

Also, the effective potential for null geodesics in our model can be calculated by using Eq. \eqref{eff2} as 
\begin{widetext}
\begin{equation}
V_{eff} = \frac{\alpha _0^2 \left(L^2 (1-c r)+2 r^2\right) \left(2 M r+q^2\right)+L^2 r^2 \left(-c r^3-2 M r+q^2+r^2\right)+2 r^6}{2 r^2 \left(\alpha _0^2 \left(2 M r+q^2\right)+r^4\right)}
\label{eff_pot}
\end{equation}
\end{widetext}
where we fix $\epsilon = 0$ for  null geodesics. Moreover, $E$ is set to be 1 for plotting purposes, as $E$ is only responsible for the shift in the effective potential amplitude and does not change the position of the photon orbit radii. The null geodesics can be obtained by solving the set of geodesic equations given by Eq. \eqref{geod_eqs_mod} numerically, and the results are shown in Fig. \ref{null_geod}. The circular photon orbits are represented by the red lines in the figure. The radii of the photon orbits correspond to the peaks of the effective potential in Fig. \ref{eff_pot_plot}.

%\begin{table*}[htb]
%\centering
%\begin{tabular}{||c|c|c||}
%\hline
%$r_p$  & $\rho_s$ & $b$      \\ \hline
%$2.54$ & $1$      & $4.5445$ \\ \hline
%$1.86$ & $2$      & $3.6484$ \\ \hline
%$1.38$ & $2.32$   & $3.1490$ \\ \hline
%\end{tabular}
%\caption{Values of the radius of circular photon path for different values of the dark matter halo density parameter $\rho_s$ and impact parameter $b = L/E$.}
%\label{tab_1}
%\end{table*}

In Fig. \ref{null_geod}, panels (a), (b), and (c) depict the behaviour of the null geodesics in the presence of varying values of the quintessence parameter \( c = 0.001, 0.0015 \), and \( 0.002 \) with $\alpha_0$ and $q$ fixed at $0.5$ and $0.4$ respectively. The unstable photon orbit, which corresponds to the peak of the effective potential, is shown by the red curve. As the magnitude of \( c \) increases up to $0.002$, we can observe noticeable variation in the geometry of the null geodesics, with a significant shift toward a repulsive behaviour, in contrast with the cases at smaller magnitudes of \( c \). The sensitivity to the parameter \( c \) is noticed in the dramatic orbital deviations resulting even from a tiny change in the value of $c$. Such a behaviour may be attributed to the inherent repulsive gravitational behaviour of a quintessence field, which, in a localised environment, could give rise to repulsive effects on the trajectories of light rays. This behaviour is theoretically expected that quintessence, as a feasible cause of repulsive gravitational effects, may modify the geodesic motion around compact objects. On the other hand, the variation of $\alpha_0$ and $q$ with $c$ fixed to $0.015$ shows the usual behaviour of null geodesics around a BH and no dramatic behaviour is seen as in the case of varying $c$. 

\subsection{Lyapunov Stability: Dynamical Systems Approach}
\begin{figure*}[tbh]
\centerline{\includegraphics[scale=0.5]{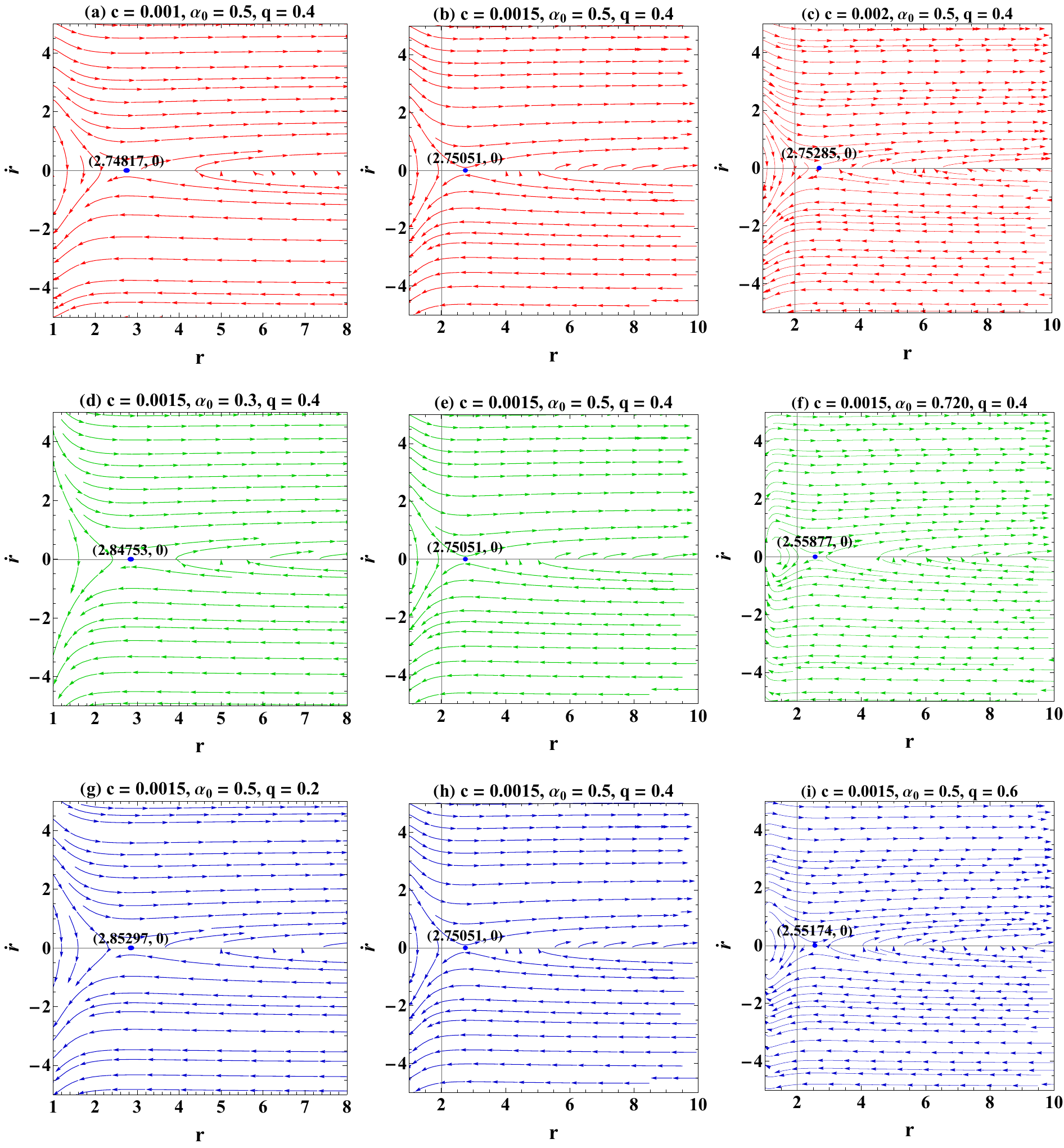}\hspace{0.1cm}}
\caption{Phase portrait of $r$ vs $\dot{r}$ for the null geodesics is shown for different combinations of the parameters $c$, $\alpha_0$, and charge $q$.}
\label{phase_plot}
\end{figure*}

To analyse the stability of null circular geodesics using the Lyapunov method, we start by constructing a dynamical system and looking at its phase space structure in the $(r, \dot{r})$ plane. For null circular geodesics, it is expected that $\dot{r} = 0$, thereby reducing the phase plane to the $(r, 0)$. By observing the phase flow dynamics in the $(r, 0)$ plane, information about the critical point that corresponds to the photon orbit radius $(r_c, 0)$ can be obtained. To do so, let us differentiate Eq. \eqref{eff1} and then eliminate $\dot{r}$ to get
\begin{equation}
\ddot{r} = - \frac{d V_{eff}}{dr},
\label{rddot}
\end{equation}
If we choose the coordinates $x_1 = \dot{r}$ and $x_2 = \dot{x_1}$, we get the following set of differential equations
\begin{widetext}
\begin{equation}
\begin{aligned}
x_1 &= \dot{r} \\
x_2 &= - \frac{d V_{eff}}{dr} \\&= \frac{1}{2 r^3 \left(\alpha _0^2 \left(2 M r+q^2\right)+r^4\right){}^2}\left[L^2 \left(r^6 \left(4 q^2-r (r (c r-2)+6 M)\right)+2 \alpha _0^2 r^3 \left(2 M \left(r^2 (2-c r)+q^2\right) +q^2 r (2-c r)\right)- \right. \right. \\& \left.\left. \hspace{4cm}\alpha _0^4 (c r-2) \left(2 M r+q^2\right)^2\right)\right].
\end{aligned}
\label{dyn_sys}
\end{equation}
\end{widetext}
The Jacobian matrix $\mathcal{J}$ of the system \eqref{dyn_sys} can be found as
\begin{equation}
\mathcal{J} = \begin{pmatrix}
0 & 1 \\
- V_{eff}''(r) & 0
\end{pmatrix}
\label{jac}
\end{equation}
 where the $''$ denotes the double differentiation with respect to $r$. The secular equation $\left|\mathcal{J} -  \lambda \mathds{I} \right| = 0$ gives the eigenvalue squared as 
 \begin{equation}
 \lambda^2 = - V_{eff} '' (r),
 \label{eig1}
\end{equation}
It can be seen that when $V_{eff}''(r) > 0$, then $\lambda^2 < 0$, which implies that the critical point represents a stable center point whereas when $V_{eff}''(r) < 0$, $\lambda^2 > 0$, which represents a saddle critical point. The phase portrait is shown in Fig. \ref{phase_plot} for all the combinations of the model parameters and quintessence parameters.
The phase portrait illustrates the phase flow of the null geodesics for different combinations of our parameters $c$, $\alpha_0$ and $q$. The circular geodesics representing the photon sphere radius are expected to be unstable at the peak of the effective potential for all of these combinations. The instability appears as any tiny disturbance to this orbit makes the photons to deviate from their circular trajectory and plummet into the BH. This scenario is illustrated in Fig. \ref{phase_plot}, where the unstable photon orbits are displayed as saddle critical points.

\subsection{Timelike Geodesics}
\begin{figure}[htb]
\centerline{\includegraphics[scale=0.5]{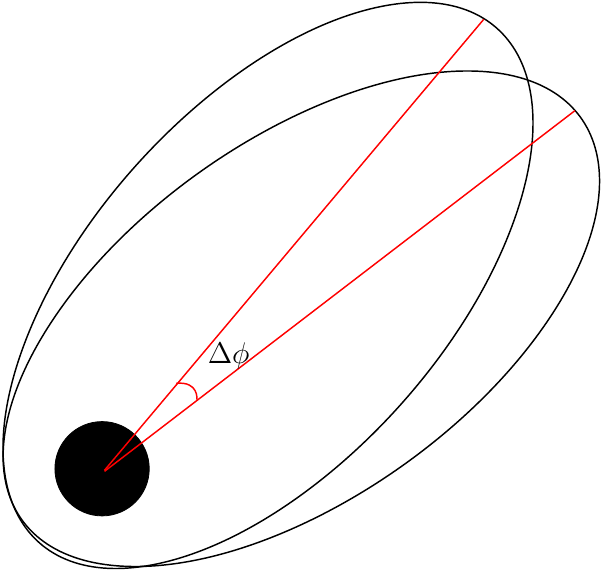}}
\caption{An illustration of the precession angle for a massive particle around the BH.}
\label{toymod}
\end{figure}

The precession of perihelion around a black hole for massive particles is a result of the described spacetime curvature in general relativity; this arises through relativistic effects, specifically those in the vicinity of the black hole. The massive particles orbiting the black hole (timelike geodesics) in the gravitational field of the BH, experience an angular deviation in the orbit through a tiny amount from being a perfect ellipse and advance the perihelion, which is the closest point of approach with each orbit. This effect is more prominent in a strong gravitational field near the black hole, and is sensitive to some extent to the black hole parameters like mass, spin, charge, as well as external fields, possibly, for instance, quintessence. This can be explained with the consideration of the timelike effective potential controlling the motion of massive particles. The turning points of the effective potential where the radial velocity momentarily equals zero give the perihelion and aphelion of the timelike orbits. It is possible that the presence of the model parameters in the metric like $\alpha_0$, charge $q$, and an external surrounding quintessence field may modify the effective potential so that orbits are no longer closed as a perfect ellipse. For each orbit, the perihelion advances. The precession angle can also be found by integrating the system of equations of motion. Metrics with different extra parameters, such as charge or quintessence, conceivably lead to various precession rates, and this might prove to be one of the significant observational probes of the BH properties and the geometry of the spacetime in the vicinity of the black hole. An illustration of the precession angle is shown in Fig. \ref{toymod}.
In terms of the new variable $u = 1/r$, the differential equation of motion governing the timelike orbits is given by \cite{Su2024Oct}
\begin{equation}
\frac{d^2 u}{d\phi^2} = - \frac{1}{2}\left(u^2 + \frac{1}{L^2}\right)\frac{dF(u)}{du} - u F(u),
\label{timelike_diff}
\end{equation}
where $F(u) = f(1/r)$. Eq. \eqref{timelike_diff} can be numerically solved for the timelike orbits of massive particles around the BH.

The effective potential for timelike orbits is given by
\begin{equation}
\begin{aligned}
V_{eff} (r) &= \frac{1}{2} \left(\frac{\left(L^2-2 r^2\right) \left(q^2-2 M r\right)}{\alpha _0^2 \left(2 M r+q^2\right)+r^4} \right. \\& \left.+ c \left(2 r^2-L^2\right) r^{-3 (w+1)} +\frac{L^2}{r^2}+3\right),
\end{aligned}
\label{effec_timelike}
\end{equation}
In the analysis of the timelike orbits in a BH spacetime, the effective potential  \( V_{\text{eff}}(r) \) can be used to predict the presence of periastron precession of timelike particles. The effective potential of timelike particles in the present model possesses a well, where there exist two turning points between which the oscillations take place repeatedly. These two turning points correspond to the points $r_{\text{min}}$ and $r_{\text{max}}$, or in other words, periastron and apastron, respectively.
These turning points are the positions at which the radial velocity $\dot{r}$ becomes zero momentarily so that the motion in this direction is repeated, and oscillatory. When combined with the angular momentum term arising from the relativistic effects, bounded oscillation takes place and, after each completed orbit, the periastron precesses gradually. Thus, the well in the effective potential not only provides for the bounded orbits but also quite naturally explains the existence of the non-zero precession angle, $\Delta \phi$, between the successive orbits. The value of \( \Delta \phi \) can be in principle, be calculated explicitly by integrating over one cycle and summing the contributions from the radial and the angular parts of the geodesic equations.
The precession of the azimuthal angle $\phi$ of the timelike orbits can be calculated from the following relation \cite{Su2024Oct}
\begin{equation}
\begin{aligned}
\Delta \phi &= \int_{r_{\text{min }}}^{r_{\text{max }}} \frac{d\phi}{dr} dr + \int_{r_{\text{max }}}^{r_{\text{min }}} \frac{d\phi}{dr} dr  - 2\pi \\&= 2 \int_{r_{\text{min }}}^{r_{\text{max }}} \frac{dr}{r^2 \sqrt{\frac{1}{b^2} - \frac{f(r)}{r^2} - \frac{f(r)}{L^2}}} - 2\pi
\end{aligned}
\label{press_angle}
\end{equation}
where $r_{\text{max}}$ and $r_{\text{min}}$ are the maximal and minimal positions of the BH at the centre (periastron and apastron).

\begin{figure*}[tbh]
\centerline{\includegraphics[scale=0.50]{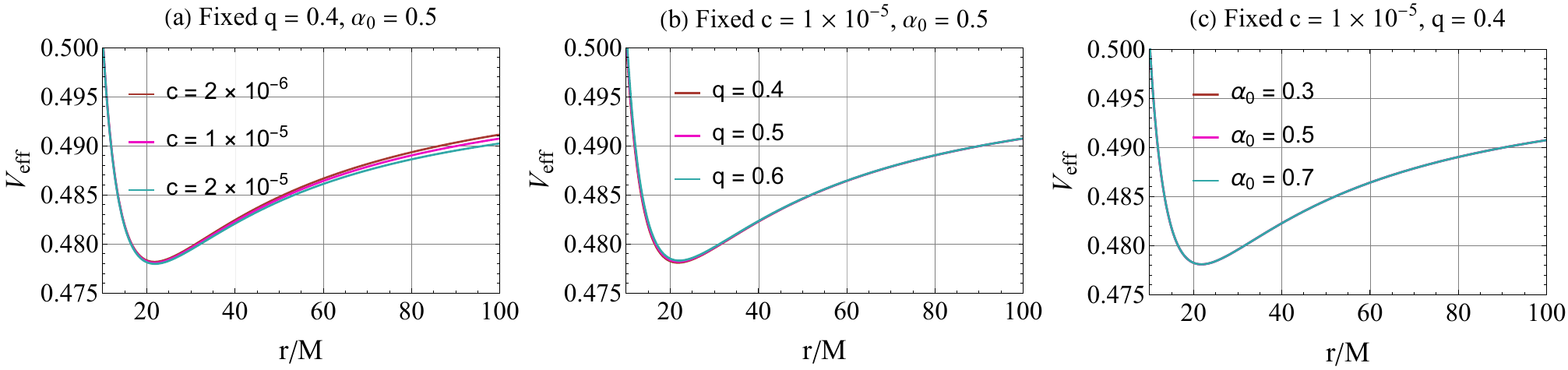}}
\caption{The plot of effective potential for timelike geodesics is shown for different combinations of the parameters $c$, $\alpha_0$, and charge $q$.}
\label{eff_pot_plot_timelike}
\end{figure*}

\begin{figure*}[!htbp]
\centerline{
\includegraphics[scale=0.35]{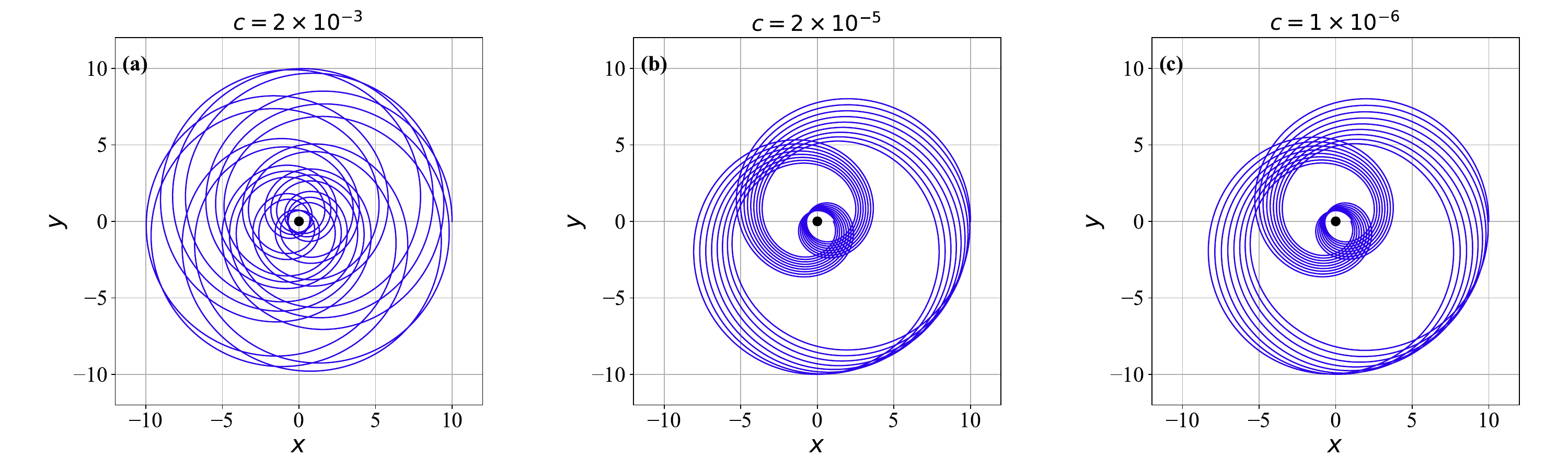}}
\caption{The timelike orbits are shown for various $c$ values setting $L = 3.2$, $\alpha_0 = 0.5$ and $q = 0.4$.}
\label{fig_precess1}
\end{figure*}

\begin{figure*}[!htbp]
\centerline{
\includegraphics[scale=0.35]{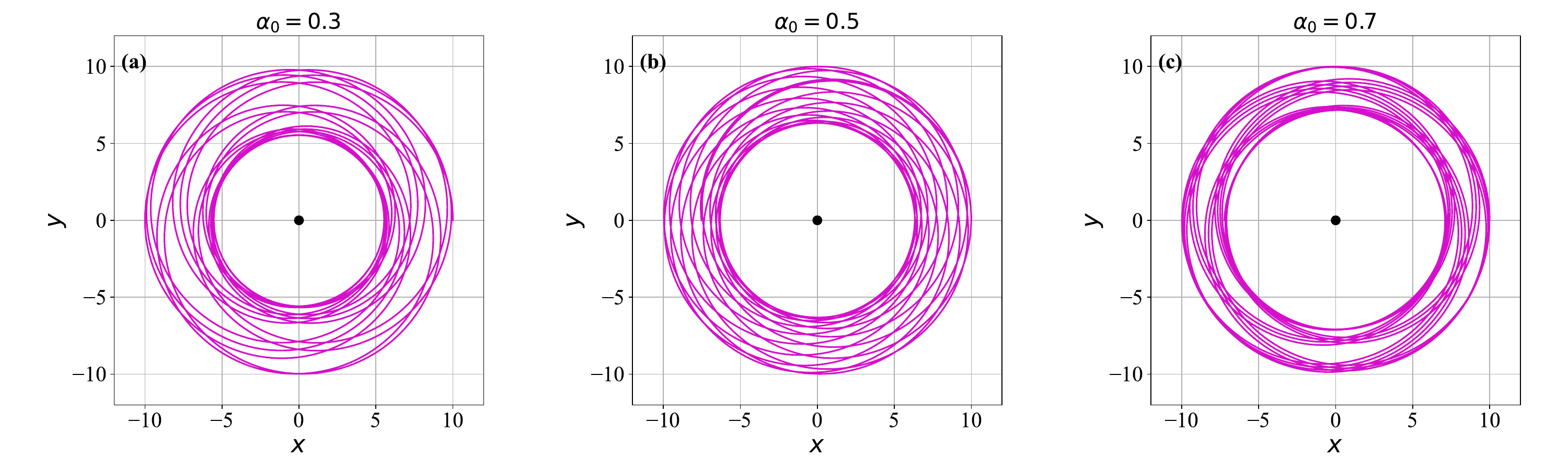}}
\caption{The timelike orbits are shown for various $\alpha_0$ values setting $c = 2 \times 10^{-6}$, $L = 3.2$ and $q = 0.4$.}
\label{fig_precess2}
\end{figure*}

\begin{figure*}[!htbp]
\centerline{
\includegraphics[scale=0.35]{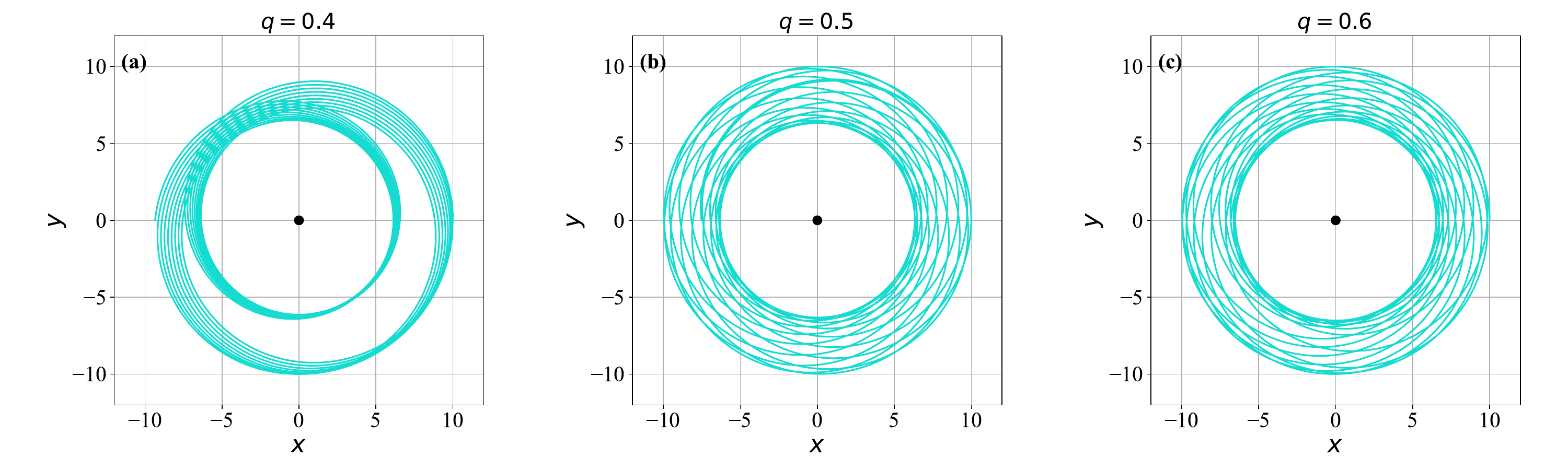}}
\caption{The timelike orbits are shown for various $q$ values setting $c = 2 \times 10^{-6}$, $\alpha_0 = 0.5$ and $L = 3.2$.}
\label{fig_precess3}
\end{figure*}

\begin{figure*}[tbh]
\centerline{\includegraphics[scale=0.4]{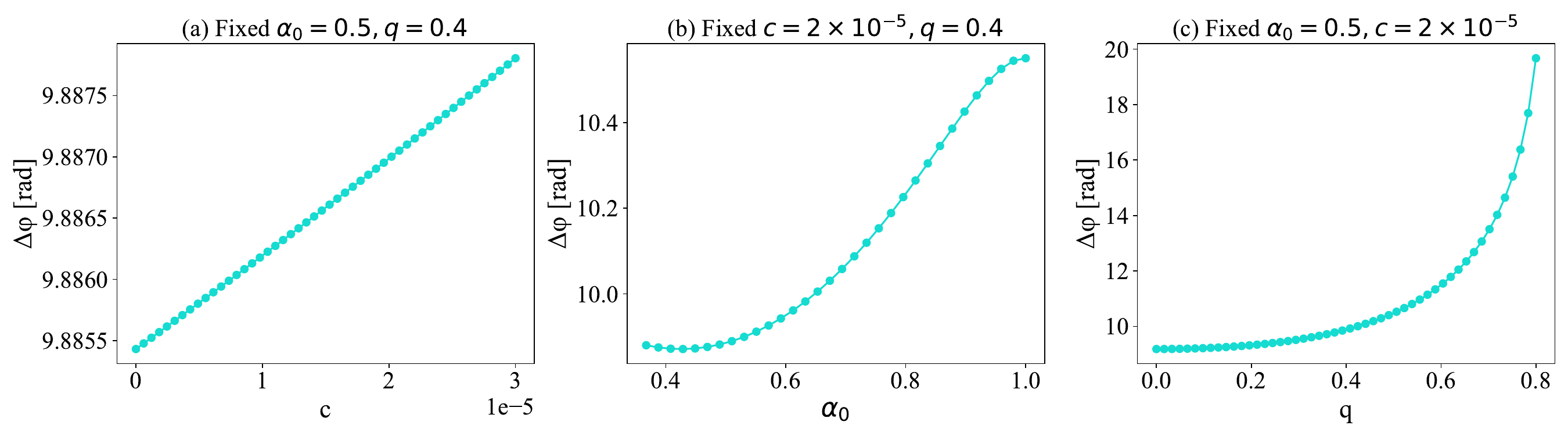}}
\caption{The plot of precession angle is shown for different combinations of the parameters $c$, $\alpha_0$, and charge $q$. Here, $L$ is set at 3.2 and energy $E$ is set at $0.95$ or in terms of impact parameter $b = L/E = 3.36$.}
\label{precess_angle}
\end{figure*}

Setting $w = -2/3$, we plot the effective potential for different combinations of $c$, $\alpha_0$ and $q$ as shown in Fig. \ref{eff_pot_plot_timelike}.
We observe that for bound orbits (depicted by wells in the effective potential) to exist, the quintessence parameter should take minimal values. The potential is observably affected by the quintessence parameter $c$, which can also be seen from Fig. \ref{fig_precess1} (a), (b) and (c). It is seen that when $c$ is larger, the orbital precession is observed very clearly, whereas with much smaller values, the orbital shapes become identical while possessing complicated orbital precessions. Such complicated orbits are elaborately discussed in Ref. \citep{Levin2008May}. Regarding the variation of massive particle orbits with $\alpha_0$, the orbital shapes are shown in Fig. \ref{fig_precess2} (a), (b) and (c). Although the effective potential is very minutely affected by $\alpha_0$, the orbital shapes are affected very strongly by tiny changes in its value. Thus, in this case, the orbital morphology is quite sensitive to small variations in the parameter $\alpha_0$. When one examines how the charge parameter affects the bound orbits, it is seen that the effective potential is very feebly affected by the parameter $q$, whereas the depth of the well increases with decreasing $q$ values. The corresponding orbital motion is shown in Fig. \ref{fig_precess3} (a), (b) and (c), which shows a noticeable difference in the orbital patterns as the value of $q$ becomes larger. At larger values, the patterns of orbits tend to take identical shapes. Finally, when the angular momentum parameter $L$ is varied, its effects on the bound orbital shapes are seen to be very significant. In this paper, we restrict ourselves to observing how the model parameters affect the shape of bound orbits qualitatively, and so the classification of these orbits is beyond the present scope.
For the understanding of the detailed classification of different patterns and orientations of orbits, one may refer to Ref. \cite{Levin2008May}. The variation of the precession angle is shown for varying values of the model parameters in Fig. \ref{precess_angle}. It can be seen that the precession angle, defined by Eq. \eqref{press_angle}, increases linearly with $c$ while non-linearly in terms of the other two parameters $\alpha_0$ and $q$.

\begin{figure*}[!htbp]
\centerline{
\includegraphics[scale=0.35]{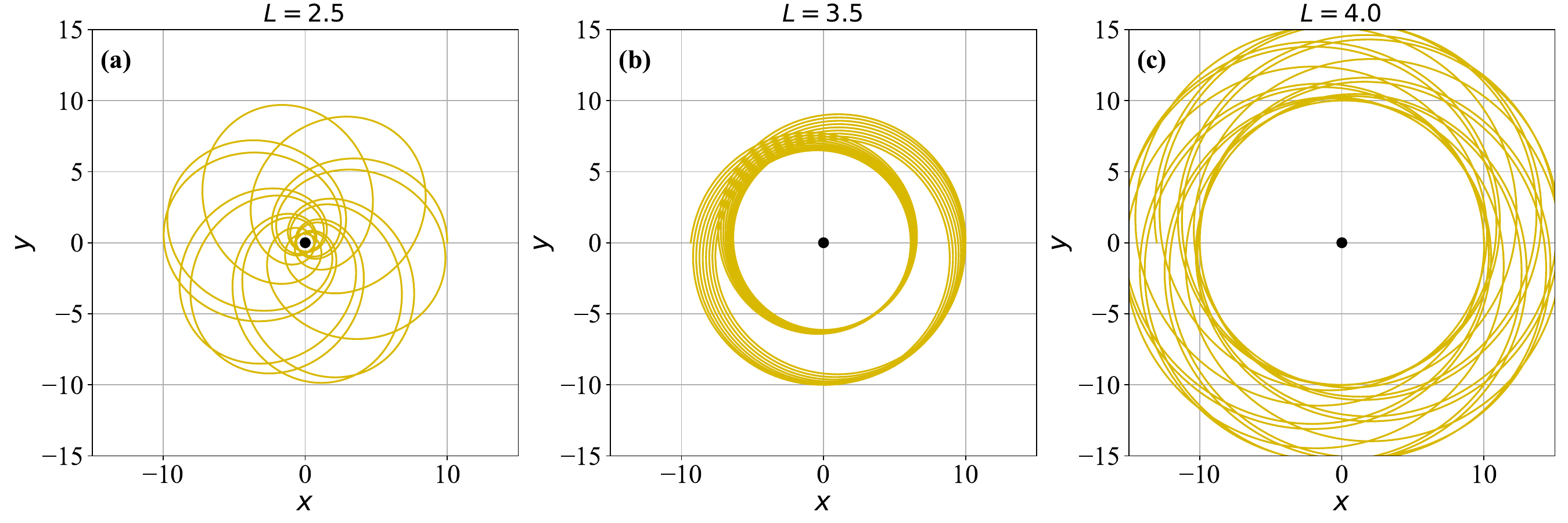}}
\caption{The timelike orbits are shown for various $L$ values setting $c = 2 \times 10^{-6}$, $\alpha_0 = 0.5$ and $q = 0.4$.}
\label{fig_precess4}
\end{figure*}

\section{Shadows}
\label{sec5}
Observational signatures of black holes are essentially characterized by shadows, providing crucial information on the geometry and spacetime of the black hole. A black hole shadow usually appears as a dark region cast by the black hole against the backlight of light from the accretion disk surrounding it or from distant sources. The size and shape of the shadow mainly depend upon the photon sphere, which is the unstable orbit of photons around the black hole, and are determined by various factors like the black hole's mass, spin, and the presence of external fields, such as quintessence or dark matter. Specifically, violations of spherical symmetry, where, for example, the rotating (Kerr) black hole introduces a distorted asymmetry that gives rise to a distorted, asymmetric shadow, opening the possibility for using shadows to distinguish between different metrics describing the appearance of black holes. Thus, the study of the shadows of black holes, as has been obtained by the Event Horizon Telescope \cite{EventHorizonTelescopeCollaboration2022May}, offers ways to verify general relativity in the strong-field regime and can even place constraints on other theories of gravity. 

In this paper, we investigate the effect of the model parameters of a Frolov BH and the quintessence parameter $c$ on the shadow radius. Before that, we need to calculate the photon orbit radius $r_p$. Following the procedure of Perlick and Tsupko \cite{Perlick2022Feb}, we can calculate the photon orbit radius through the following relation
\begin{equation}
f'(r_p) r_p - 2f(r_p) = 0,
\label{phot_orb_rad_relation}
\end{equation}
which gives
\begin{equation}
\begin{aligned}
&
r_p \left[-c (-3 w-1) r_p^{-3 w-2}-\frac{2 M r_p^2}{\alpha _0^2 \left(2 M r_p+q^2\right)+r_p^4} \right. \\& \left.+\frac{r_p^2 \left(2 M r_p-q^2\right) \left(2 \alpha _0^2 M+4 r_p^3\right)}{\left(\alpha _0^2 \left(2 M r_p+q^2\right)+r_p^4\right)^2}-\frac{2 r_p \left(2 M r_p-q^2\right)}{\alpha _0^2 \left(2 M r_p+q^2\right)+r_p^4}\right] \\ &  -2 \left[-c r_p^{-3 w-1}-\frac{r_p^2 \left(2 M r_p-q^2\right)}{\alpha _0^2 \left(2 M r_p+q^2\right)+r_p^4}+1\right] =0
\end{aligned}
\label{phot_orb_cond_eq}
\end{equation}
Eq. \eqref{phot_orb_cond_eq} is a complicated relation that cannot be solved analytically. To solve this equation, we resort to a numerical approach to obtain the photon orbit radii $r_p$. After obtaining the photon orbit radii, 
we may calculate the shadow radii directly using the relation \cite{Perlick2022Feb}
\begin{equation}
r_s = \frac{r_p}{\sqrt{f(r)}\left. \right|_{r_p}}
\label{shad_rad_main}
\end{equation}
\begin{figure*}[tbh]
\centerline{\includegraphics[scale=0.35]{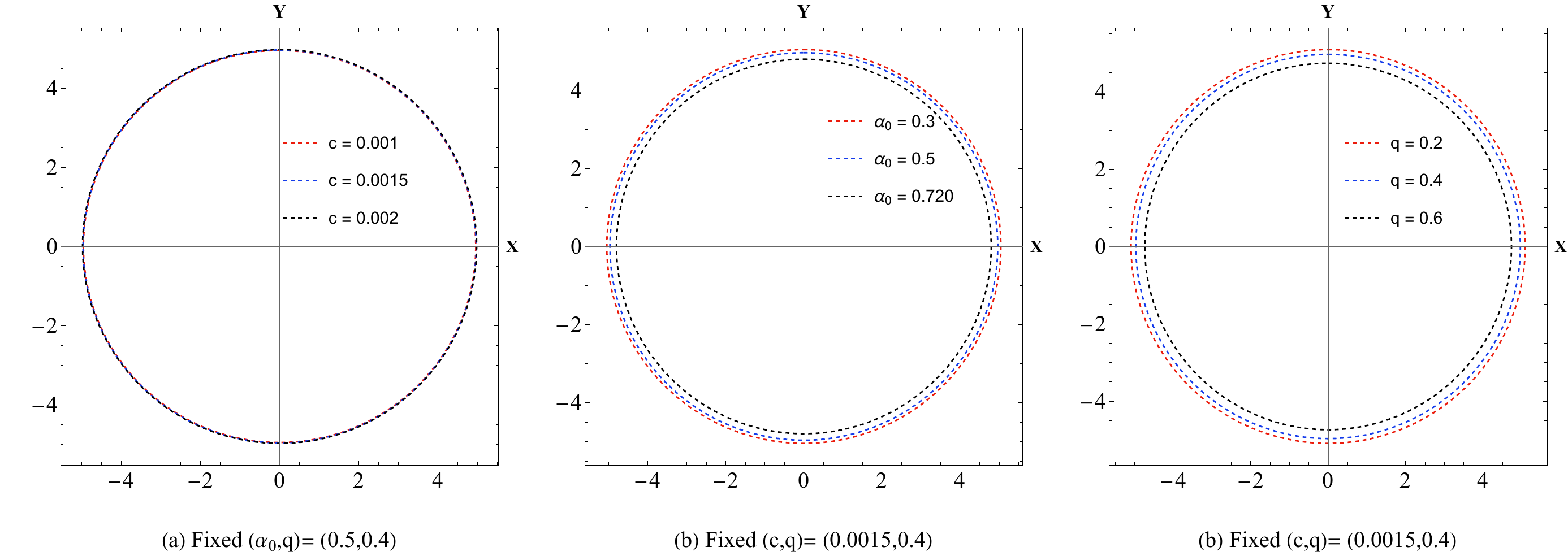}}
\vspace{0.5cm}
\centerline{\includegraphics[scale=0.6]{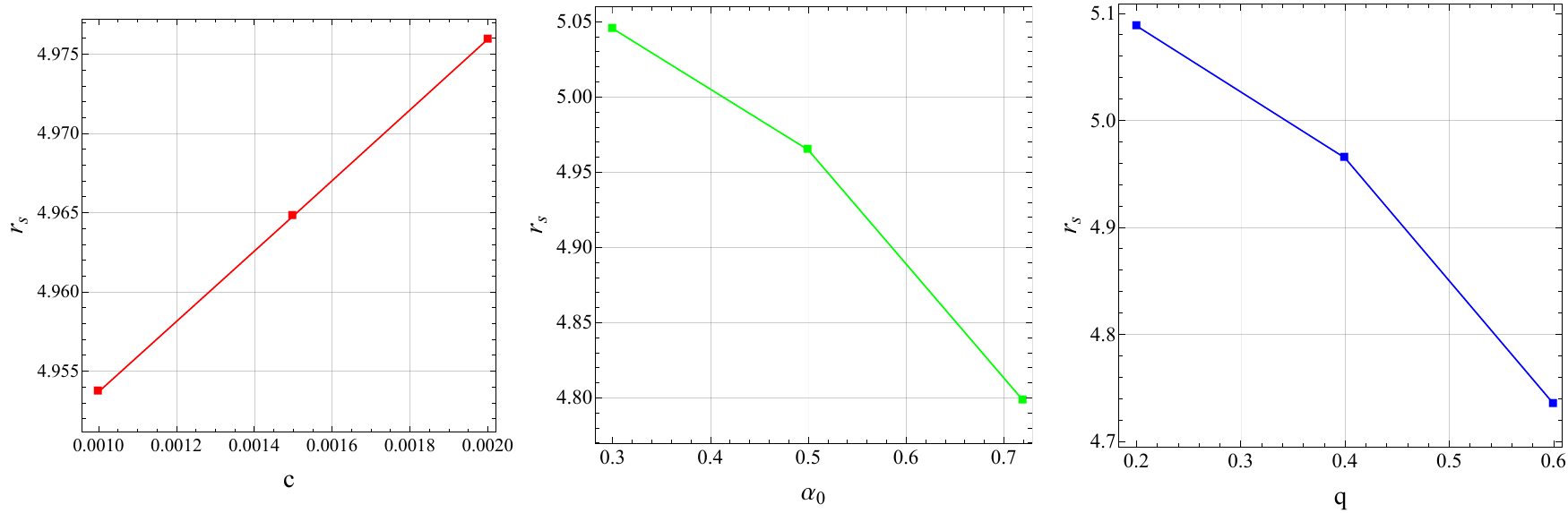}}
\caption{The plot of the shadows (upper panel) and the variation of shadow radius (lower panel) is shown for different combinations of the parameters $c$, $\alpha_0$, and charge $q$.}
\label{shadow_plot}
\end{figure*}

To obtain the actual shadow image in the observer's frame, one may introduce the celestial coordinates $X$ and $Y$ as \cite{Perlick2022Feb} 
\begin{equation}
\begin{aligned}
X &= \lim_{r_o \to \infty} \left( -r_o^2 \sin \theta_o \frac{d\phi}{dr}\right),\\
Y &= \lim_{r_o \to \infty}\left(r_o^2 \frac{d\theta}{dr} \right).
\end{aligned}
\label{xandy}
\end{equation}
with $r_o$ as the observer's location and $\theta_o$ as the angle of inclination. 
Fig. \ref{shadow_plot} shows the shadow behaviour affected by the model parameters. Evidently, the quintessence parameter $c$ does not seem to affect the shadow radii drastically. However, an observable effect is seen with the variation of the parameter $\alpha_0$ and charge $q$. The shadow radius increases with an increase in the quintessence parameter $c$, while it decreases with an increase in the values of $\alpha_0$ and $q$. This is shown in the lower panel of Fig. \ref{shadow_plot}.
\begin{table*}[!htb]
\centering
\begin{tabular}{||c|c|c|c||}
\hline
%\rowcolor[]{} 
$\mathbf{(\alpha_0, q )}$           & $\mathbf{c}$        & $\mathbf{r_{p}}$ & $\mathbf{r_{s}}$ \\ \hline
%\rowcolor[]{} 
                                     & 0.001               & 2.74812          & 4.95368          \\ \cline{2-4} 
%\rowcolor[]{} 
$(0.5, 0.4)$ & 0.0015              & 2.75051          & 4.96477          \\ \cline{2-4} 
%\rowcolor[]{} 
                                     & 0.002               & 2.75285          & 4.97593          \\ \hline
%\rowcolor[]{} 
$\mathbf{(c, q)}$                   & $\mathbf{\alpha_0}$ & $\mathbf{r_{p}}$ & $\mathbf{r_{s}}$ \\ \hline
                                     & 0.3                 & 2.84753          & 5.04540          \\ \cline{2-4} 
$(0.0015, 0.4)$                      & 0.5                 & 2.75051          & 4.96477          \\ \cline{2-4} 
                                     & 0.720               & 2.55877          & 4.79807          \\ \hline
%\rowcolor[]{} 
$\mathbf{(\alpha_0 , c )}$          & $\mathbf{q}$        & $\mathbf{r_{p}}$ & $\mathbf{r_{s}}$ \\ \hline
                                     & 0.2                 & 2.85297          & 5.08823          \\ \cline{2-4} 
$(0.5, 0.0015)$                      & 0.4                 & 2.75051          & 4.96477          \\ \cline{2-4} 
                                     & 0.6                 & 2.55174          & 4.73505          \\ \hline
\end{tabular}
\caption{The values of photon orbit radii and the shadow radii for different combination of the parameters $c$, $\alpha_0$, and charge $q$.}
\label{Tab_phot_rad_shad}
\end{table*}

\subsection{Parameter constraints}
The EHT's latest horizon-scale image of Sgr A$^*$ \cite{EventHorizonTelescopeCollaboration2022May} provides the ideal opportunity to investigate gravity and fundamental physics in the strong-gravity regime. Sgr A$^*$'s proximity to us allows it to be much simpler to estimate its mass and distance, and thus its mass-to-distance ratio. This is a major advantage over M87*, whose mass is not tightly constrained based on the stellar dynamics observations. The Sgr A$^*$'s mass ($O(10^6)M_\odot$) is significantly lower than the M87$^*$'s mass ($O(10^9)M_\odot$), providing a window to look into fundamental physics in a strong curvature domain in a completely new way. With the same justification, the Sgr A's shadow can provide more stringent constraints compared to the M87's on fundamental parameters as well \cite{Vagnozzi2023Jul}.

The EHT measured the fractional deviation between the estimated shadow radius $r_s$ and the shadow radius of a Schwarzschild BH $r_{sch} = 3\sqrt{3} M$, which is defined as
\begin{equation}
\delta \equiv \frac{r_s}{r_{sch}} - 1 = \frac{r_{s}}{3\sqrt{3} M} - 1.
\label{del_shad}
\end{equation}
The Keck and VLTI measurements gave the following constrains on $\delta$ \cite{Do2019Jul,Abuter2020Apr,Vagnozzi2023Jul}
\[ \delta = -0.04^{+0.09}_{-0.10} \text{  (Keck) \quad  and  \quad } 
\delta = -0.08^{+0.09}_{-0.09} \text{  (VLTI)} \]
For an easier analysis, one may calculate the average value of $\delta$ obtained from the Keck and VLTI instruments, as these values are uncorrelated since they come from independent setups. The averaged value of $\delta$ is
\begin{equation}
\delta \approx -0.060 \pm 0.065.
\label{av_del}
\end{equation}
By assuming a Gaussian distribution of the posterior shapes, the $1\sigma$ and $2\sigma$ confidence intervals for $\delta$ are
\begin{equation}
 -0.125 \lesssim \delta  \lesssim 0.005 \quad (1 \sigma) \text{  and }  -0.19 \lesssim \delta  \lesssim 0.07 \quad(2 \sigma).
 \label{const_aver}
\end{equation}
Extracting $r_s$ from Eq. \eqref{del_shad}, we get 
\begin{equation}
\frac{r_s}{M} = 3 \sqrt{3} (\delta + 1),
\label{rs_m}
\end{equation}
which, when combined with the constraints Eq. \eqref{const_aver}, gives
\begin{equation}
4.55 \lesssim r_s/M \lesssim 5.22, \,(1\sigma) \text{ and } 
4.21 \lesssim r_s/M \lesssim 5.56, \,(2 \sigma).
\label{rs_const}
\end{equation}

\begin{figure*}[htb]
\centerline{\includegraphics[scale=0.3]{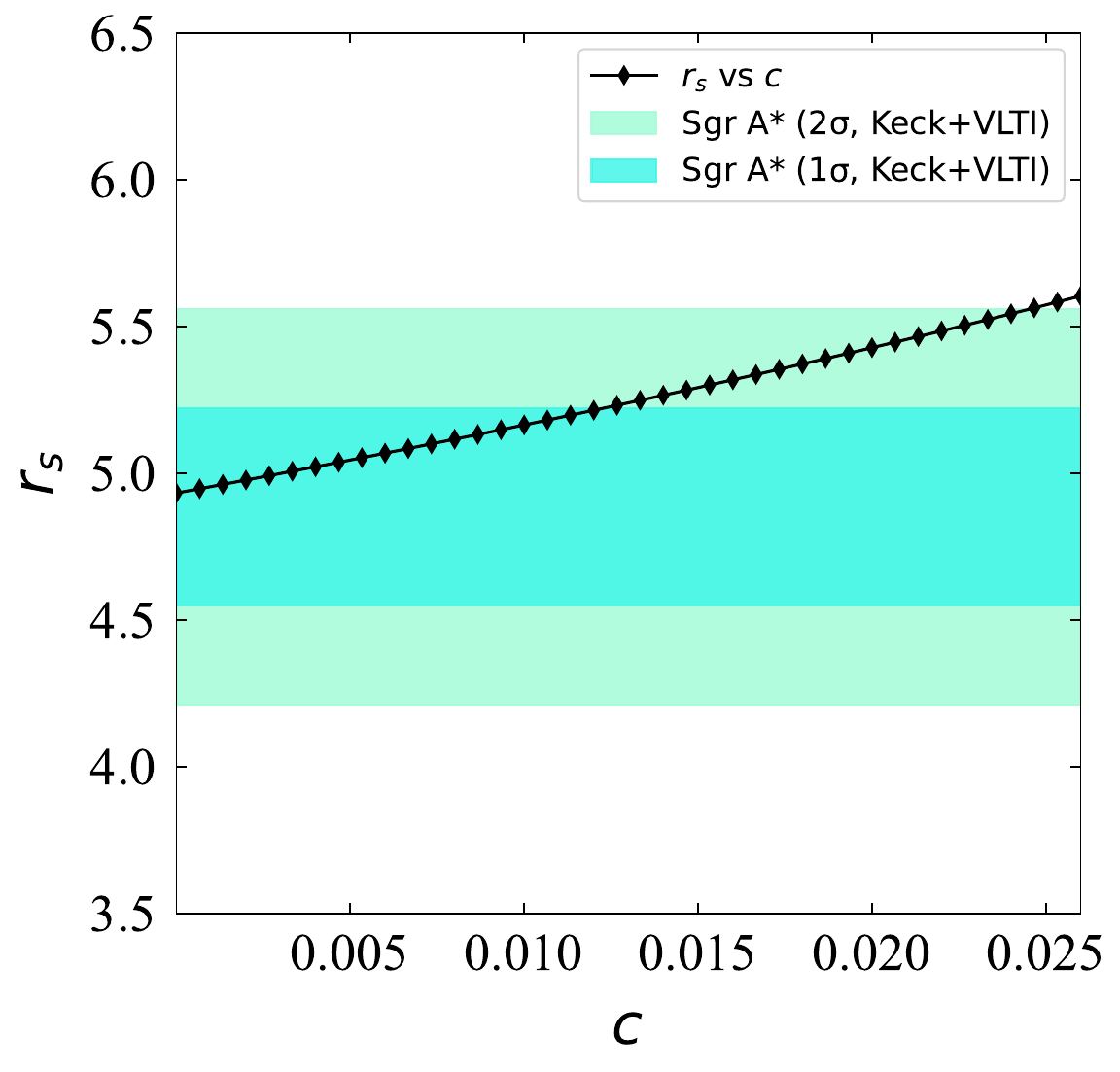}\hspace{0.2cm}\includegraphics[scale=0.3]{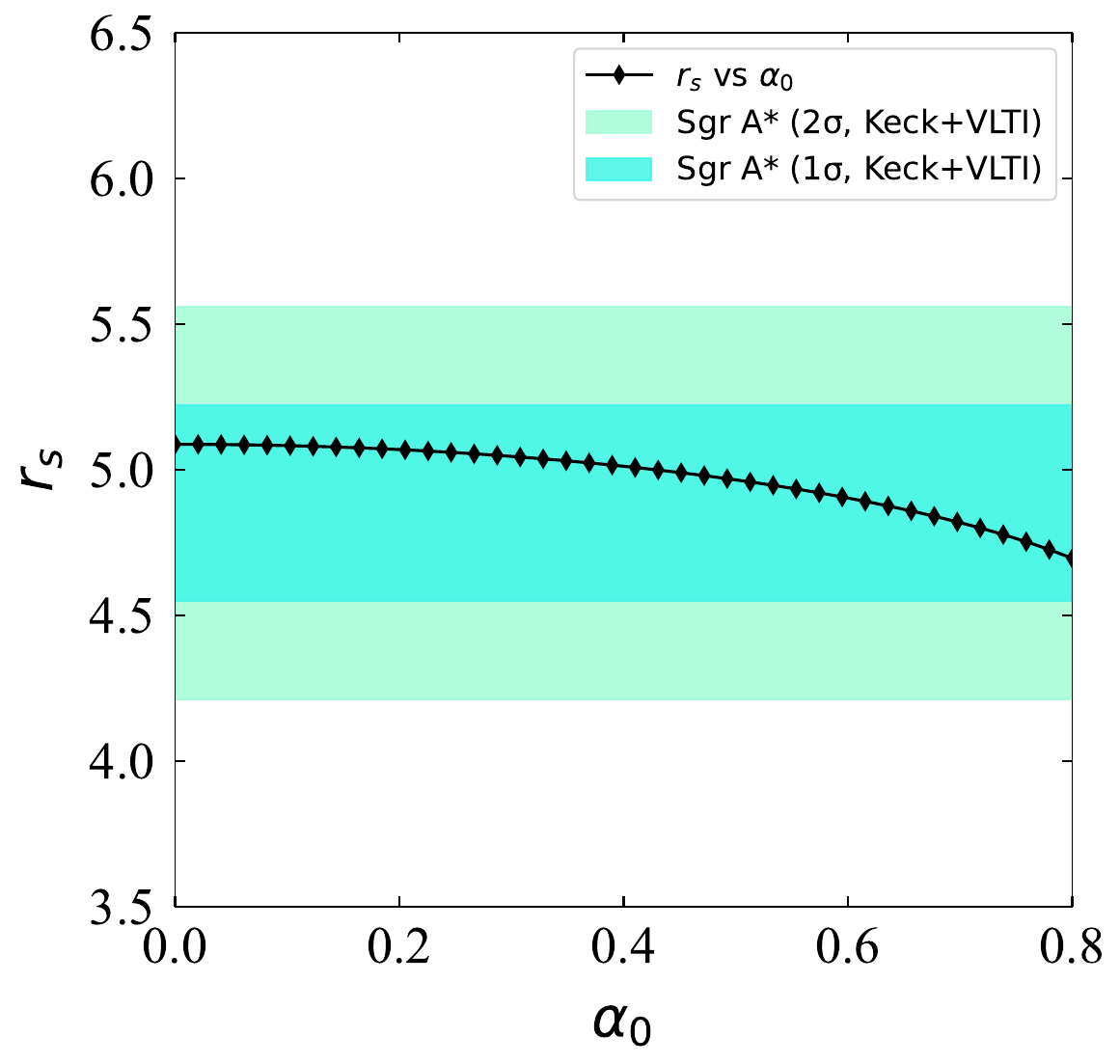}\hspace{0.2cm}\includegraphics[scale=0.3]{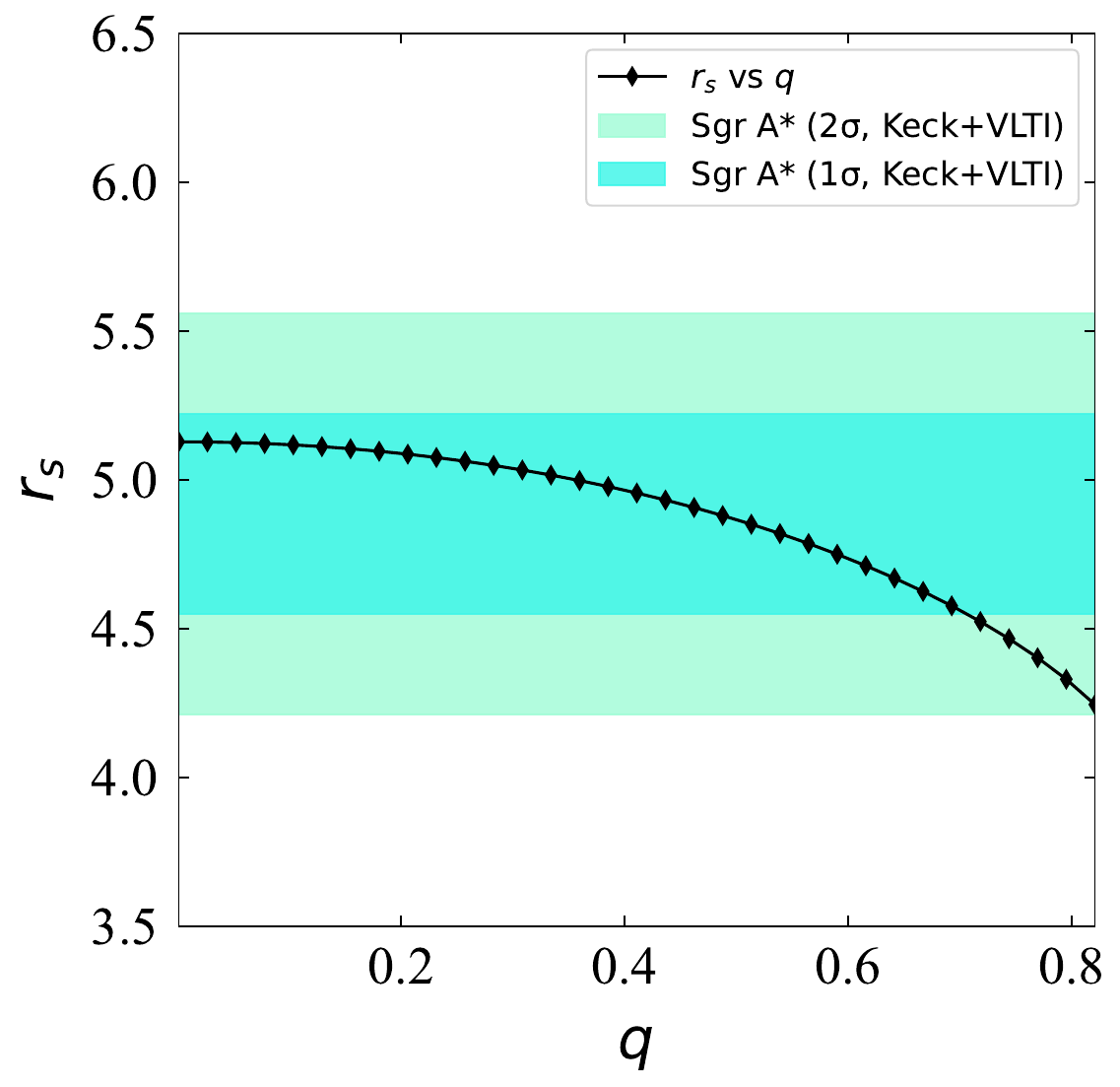}}
\centerline{\hspace{3cm} (a) \hfill \hspace{0.3cm}(b) \hfill (c) \hspace{2.2cm}}
\caption{Variation of shadow radius with the model parameters $c$, $\alpha_0$, and charge $q$ in comparison with observed shadow radii by Keck/VLTI of Sgr $A^*$ BH. In (a), $\alpha_0$ and $q$ are fixed to 0.5 and 0.4, in (b) $c$ and $q$ are fixed at $0.0015$ and $0.4$ and in (c) $\alpha_0$ and $c$ are fixed at $0.5$ and $0.0015$, respectively. }
\label{shad_obs_plot}
\end{figure*}

Now, using Eq. \eqref{shad_rad_main}, Eq. \eqref{rs_m} and the confidence intervals \eqref{rs_const}, we plot $r_s$ with respect to our model parameters $c$, $\alpha_0$ and $q$, which is shown in the Fig. \ref{shad_obs_plot}. In section \ref{sec2}, it was already mentioned about the allowed ranges of the parameter $\alpha_0$ and $q$, which are verified in Fig. \ref{shad_obs_plot} (b) and (c), that both lie within both $1\sigma$ and $2\sigma$ confidence levels. The free parameter of our model, which is the quintessence parameter $c$ is found to evolve well within the $1\sigma$ and $2\sigma$ confidence regions as shown in Fig. \ref{shad_obs_plot} (a). The constraints set from the analysis on $c$ are found to be \begin{equation}
\begin{aligned}
&0 \lesssim \, c \, \lesssim 0.01201 \quad (1\sigma) \\
&0 \lesssim \, c \, \lesssim 0.02467 \quad (2\sigma) 
\end{aligned}
\label{const_c}
\end{equation}

\section{Conclusion}
\label{conc}
In this paper, we considered the possibility of a Frolov BH surrounded by a quintessence field. In the first part, we looked at the thermodynamic parameters of the BH system, particularly the Hawking temperature, specific heat, and free energy. The results suggest that the BH is locally stable at lower horizon radii but locally unstable at higher horizon radii. The decreasing value of the quintessence parameter shifts the Davies point towards higher horizon radii, however contrasting behaviour is seen with the Hubble length parameter $\alpha_0$ and charge $q$. The free energy is seen to remain positive for all permissible horizon radii, meaning that the BH is globally thermodynamically unstable.

In the second part, we derive the effective potential of the BH system and the geodesic equations resulting from our Frolov + quintessence BH solution. The radii of circular orbits of light rays increase with increasing values of the quintessence parameter $c$. This is further demonstrated numerically by using backward ray-tracing of light rays around the BH. It is expected that the unstable photon orbits should correspond to saddle points in the $(r, \dot{r})$ phase plane, confirming the unstable. One important result of this analysis is that the quintessence parameter drastically affects the null geodesics in the sense that high value of quintessence surrounding the BH overpowers the curvature effects of the BH, thereby resisting the inflow of light rays into the BH. Next, we studied the nature of timelike geodesics for massive particles around our BH spacetime. We numerically solved the timelike geodesic equations and obtained the precession angles of massive particles around the BH spacetime. It is found that the quintessence parameter $c$, noticeably affect the precession pattern at higher values and the precession angles increases linearly with increasing quintessence effects. The parameters $\alpha_0$ and $q$ seem to impact the precession patterns noticeably, but in this case the precession angles varies non-linearly with increasing values of $\alpha_0$ and $q$.  Finally, we analysed the shadow cast by the Frolov BH in the presence of quintessence and further constrained the model parameters for the observational shadow data.

\bibliography{bibliography.bib}
\end{document}